\documentclass[preprint, 12pt, authoryear]{elsarticle}

\usepackage{aliases}
\usepackage{comments}

\journal{Computer Methods in Applied Mechanics and Engineering}

\begin{document}

\begin{frontmatter}



\title{Multi-fidelity Hamiltonian Monte Carlo}






\author[stanford]{Dhruv V. Patel}
\author[hawaii]{Jonghyun Lee}
\author[erdc]{Matthew W. Farthing}
\author[stanford_civil]{Peter K. Kitanidis}
\author[stanford]{Eric F. Darve}

\affiliation[stanford]{organization={Department of Mechanical Engineering},
            addressline={Stanford University}, 
            city={Stanford},
            state={CA},
            country={USA}}

\affiliation[hawaii]{organization={Department of Civil and Environmental Engineering},
            addressline={University of Hawai`i at Manoa}, 
            city={Honolulu},
            state={HI},
            country={USA}}

\affiliation[erdc]{organization={U.S. Army Engineering Research and Development Center},
            city={Vicksburg},
            state={MS},
            country={USA}}  

\affiliation[stanford_civil]{organization={Department of Civil and Environmental Engineering},
            addressline={Stanford University}, 
            city={Stanford},
            state={CA},
            country={USA}}

\begin{abstract}
Numerous applications in biology, statistics, science, and engineering require generating samples from complex, high-dimensional probability distributions. In recent years, the Hamiltonian Monte Carlo (HMC) method has emerged as a state-of-the-art Markov chain Monte Carlo (MCMC) technique, exploiting the shape of such high-dimensional target distributions to efficiently generate samples. Despite its impressive empirical success and increasing popularity, its wide-scale adoption remains limited due to the high computational cost of gradient calculation. Moreover, applying this method is impossible when the gradient of the posterior cannot be computed (for example, with black-box forward model simulators and/or with non-differentiable priors). To overcome these challenges, we propose a novel two-stage Hamiltonian Monte Carlo algorithm with a surrogate model that supports cost-effective gradient computation. In this multi-fidelity algorithm, the acceptance probability is computed in the first stage via a standard HMC proposal using an inexpensive differentiable surrogate model (which can be based on deep learning (DL)-based surrogate), and, if the proposal is accepted, the posterior is evaluated in the second stage using the high-fidelity (HF) numerical solver. Splitting the standard HMC algorithm into these two stages allows for approximating the gradient of the posterior efficiently (thus retaining advantages of HMC, such as scalability to high dimensions and faster convergence), while producing accurate posterior samples by using HF numerical solvers in the second stage. We demonstrate the effectiveness of this algorithm for a range of problems, including linear and nonlinear Bayesian inverse problems with \textit{in-silico} data and a nonlinear hydraulic tomography problem using experimental data. The proposed algorithm is shown to seamlessly integrate with various low-fidelity (LF) and HF models, priors, and datasets, highlighting its broad versatility and practical applicability. Remarkably, our proposed method outperforms the traditional HMC algorithm in both computational and statistical efficiency by several orders of magnitude, all while retaining or improving the accuracy in computed posterior statistics. This suggests an enticing potential for its adoption as a viable substitute for HMC, even within a white-box setting.
\end{abstract}

\begin{keyword}
Hamiltonian Monte Carlo \sep Multi-fidelity modeling \sep Uncertainty quantification \sep Bayesian inference \sep Inverse problems
\end{keyword}

\end{frontmatter}


\section{Introduction}
\label{sec:intro}
The ability to sample from a probability distribution (which might be known only up to a normalizing constant) has wide applications in science and engineering. This capability is of paramount importance in biology for generating equilibrium configurations of bio-molecules \citep{Boomsma2013PHAISTOSAF, habeck2005replica}.  In chemistry, sampling techniques are integral to molecular dynamics simulations and the estimation of chemical reaction rates, crucial for advancing drug and material development \citep{carter1989constrained, rosso2002use, zheng2013rapid}.  Sampling methods are also instrumental for computing high-dimensional integrals in statistics \citep{Gelfand1990, Brooks2011} and machine learning \citep{Andrieu2003}, for counting and volume computation \citep{dyer1991random}, and for deep generative modeling tasks \citep{koller2009probabilistic, Nijkamp2019, ho2020denoising}. In finance, sampling is used to find the expected return of portfolio \citep{detemple2003monte}. Sampling is also key for developing robust optimizers to escape local minima/saddle points and avoid overfitting \citep{welling2011bayesian, dauphin2014identifying}. Lastly, sampling methodologies are indispensable for solving Bayesian inverse problems \citep{Dashti2017, Martin2012, kaipio2006statistical}.

Markov Chain Monte Carlo (MCMC) is a popular method for generating samples from unnormalized probability distributions \citep{Gelman2014}. The strength of this method lies in the fact that it can produce unbiased samples with convergence guarantees for the Quantities of Interest (QoIs) with minimal requirements on the target distribution. There exist a variety of MCMC algorithms, such as random-walk Metropolis \citep{Metropolis1953}, Gibbs sampling \citep{geman1984stochastic}, adaptive Metropolis Hastings \citep{Haario2001}. All of these algorithms come with convergence guarantees, but this guarantee comes at a price of extremely slow convergence---exploring all the relevant parts of the parameter space that has non-negligible probability mass under the given distribution may take an unacceptably long time, as these vanilla MCMC methods typically proceed by taking random jumps around the current position without taking into account the shape of the target distribution. 

Over the years, a variety of methods have been developed to overcome the challenges associated with slow convergence. Many of these methods~\citep{roberts1996exponential, girolami2011riemann, Martin2012} achieve faster convergence by exploiting the geometry of the target probability density. In this manuscript, we limit ourselves to one such method---the Hamiltonian (or Hybrid) Monte Carlo (HMC) method~\citep{duane1987hybrid, Gelman2014}, which is a state-of-the-art MCMC method that utilizes the gradient information of the target probability density to improve the overall convergence rate. HMC is able to suppress the random walk behavior of vanilla MCMC methods by the clever trick of introducing an auxiliary variable scheme that transforms the problem of sampling from a target distribution into the problem of simulating Hamiltonian dynamics. HMC has shown tremendous promise in a variety of application domains due to its ability to characterize \textit{high-dimensional} probability densities at \textit{a faster convergence} rate.  The cost of HMC per independent sample from a target density of dimension D is roughly $\mathcal{O}(D^{5/4})$, which stands in sharp contrast with the $\mathcal{O}(D^2)$ cost of random-walk Metropolis \citep{creutz1988global, hoffman2014no}.

Despite its promising features, several hindrances still impede the widespread adoption of HMC for science and engineering problems. Firstly, a major barrier to its universal adoption is the stringent requirement of necessitating the gradient of the target probability density. For most practical problems of interest, the analytical formula for this gradient is unavailable, mandating its numerical computation. However, computing such a numerical gradient is often impossible in many practical settings, such as those involving black-box forward model simulators and/or non-differentiable priors. In such scenarios, practitioners must resort to other inefficient MCMC methods that do not require gradient computations. Secondly, even in application settings where the gradient can be computed with reasonable accuracy (i.e., in a white-box setting), the actual gradient computation can be expensive. This is particularly true for problems governed by partial differential equations (PDEs), (like the ones considered in this manuscript). For such problems, evaluating one gradient typically entails solving two PDE problems (a forward and an adjoint problem) \citep{plessix2006}, rendering the algorithm computationally demanding. As HMC is often run for many steps (on the order of $10^3$--$10^5$), and each step requires two PDE solves, the computational burden can be significant. Finally, the overall efficiency of HMC is highly sensitive to its hyperparameters: the step size ($\epsilon$) and the number of leapfrog steps ($L$). Tuning these parameters for better statistical efficiency (higher effective sample size (ESS)) often results in a degradation of computational efficiency~\citep{betancourt2017conceptual}, as it requires a greater number of samples (i.e., more number of expensive forward and adjoint solves). For example, while a longer trajectory length ($=\epsilon \times L$) theoretically yields uncorrelated samples and thus improved statistical efficiency (i.e., higher ESS), however, it typically comes with a lower acceptance rate for many practical problems, necessitating more evaluations of the target density (and its gradient), thereby decreasing overall computational efficiency.

To overcome these challenges in this manuscript, we propose a novel Multi-Fidelity Hamiltonian Monte Carlo (MFHMC) algorithm, which leverages a surrogate forward model with an easy-to-compute gradient along with a high-fidelity (HF) numerical solver. This two-stage algorithm enables efficient and accurate sampling while being significantly computationally less expensive. While the algorithm proposed in this manuscript is valid for any application of generating samples from an un-normalized probability distribution, here we focus on the problem of generating samples from the posterior density of a Bayesian inverse problem, where the forward model is typically defined by a PDE. Within this context, the method proposed here proceeds in two steps:
\begin{enumerate}
    \item In the first step, a surrogate model is developed for the forward map (mapping parameter to solution field). One can use any surrogate model, such as a proper orthogonal decomposition (POD)-based surrogate or Deep Neural Network (DNN)-based surrogate model for this \textit{offline} step. The only criteria for this low-fidelity (LF) surrogate model is its gradient should be computable cheaply. However, it does not necessarily have to be very accurate.
    \item In the second step, this surrogate model is used in the proposed MFHMC algorithm to generate samples from the target posterior density. This algorithm, which is the novel contribution of this manuscript, is described briefly below.
\end{enumerate}

The key idea of the proposed multi-fidelity algorithm (outlined in \cref{fig:workflow}) is to split the standard HMC algorithm into two stages: 
\begin{figure}[htbp]
    \centering \includegraphics[width=0.65\linewidth]{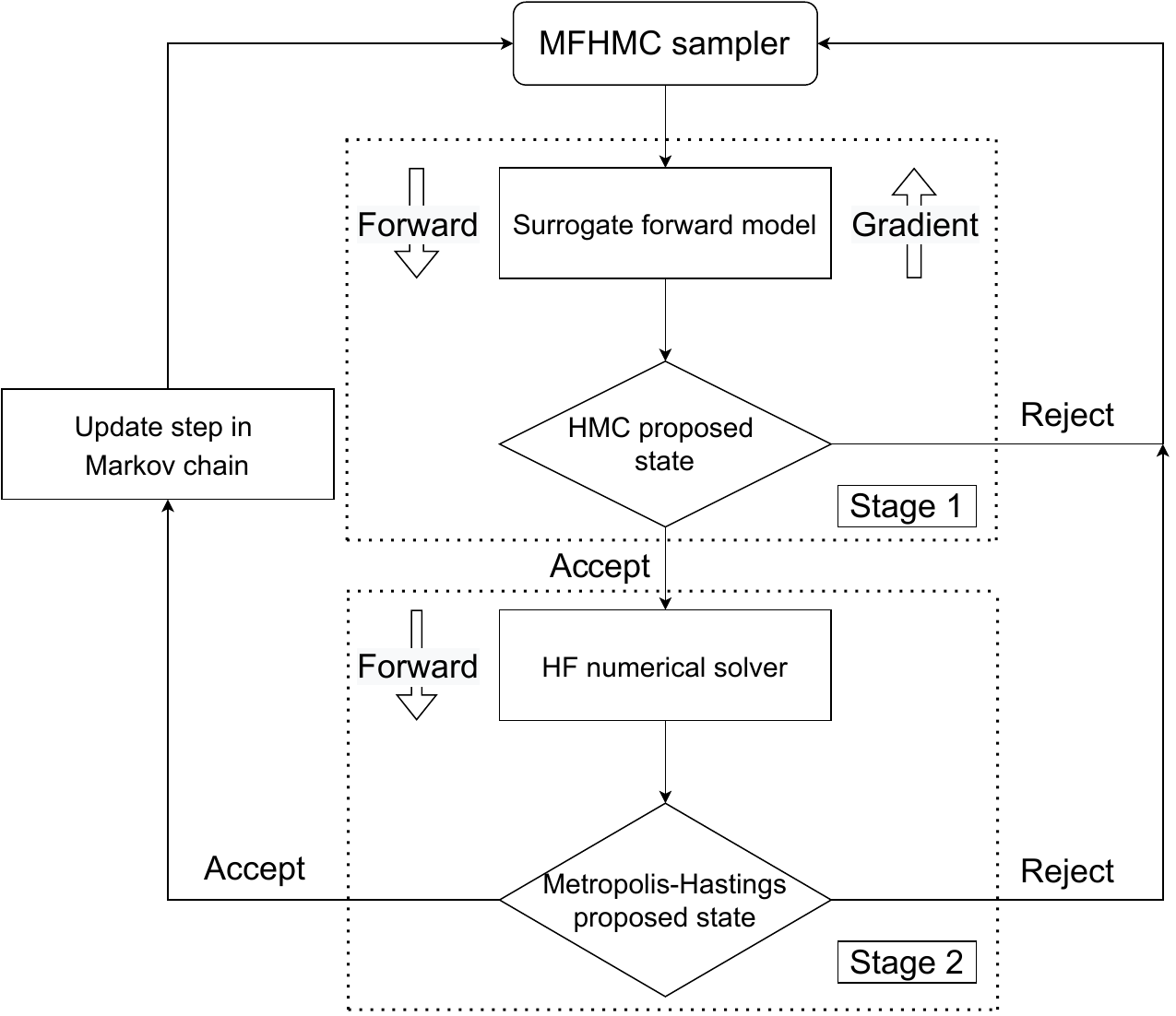}
    \caption{Outline of the proposed two-stage MFHMC algorithm: In the first stage of the algorithm a surrogate forward model (with an easy-to-compute gradient) is used in the standard HMC step. If a given sample is accepted by this first stage, then it is passed to the second stage, where a high-fidelity numerical solver is used in the Metropolis-Hastings step (for which only a forward model evaluation is required with no gradient requirement) to produce accurate posterior samples. If a sample is rejected in either stages, we stay at the current Markov chain state.}
    \label{fig:workflow}
\end{figure}
\begin{enumerate}[{Stage} 1.]
    \item In the first stage, the surrogate forward model is used (in the likelihood term) to propose samples following the standard HMC proposal. Since this surrogate model is selected in such a way that its gradient can be computed cheaply, it allows the use of the HMC algorithm (which in turn enables scalability to high-dimensional parameter space and a faster convergence rate). DNN-based models are a particularly appealing choice for such surrogate models, as (by leveraging automatic differentiation capabilities of modern machine learning libraries) it allows for computing gradients accurately at \textit{almost no additional cost}.
    \item The samples accepted by the first stage are passed to the second stage, where the HF numerical simulator is used in the Metropolis-Hastings (MH) MCMC step. The acceptance ratio for this stage is modified to take into account the first stage. Since the MH step does not require any gradient, it opens up the opportunity of using HF \textit{black-box simulators} in this stage to correct errors introduced by the LF surrogate model in the first stage and produce accurate posterior samples as output.
\end{enumerate}

We note that two-stage MCMC algorithms have been proposed in the past \citep{christen2005markov, Efendiev2006}. However, these algorithms were developed for random walk Metropolis-Hastings MCMC algorithms, which fail to scale to high-dimensional parameter space, thus limiting their practical utility. Unlike those works, this manuscript proposes a two-stage algorithm for HMC, the current state-of-the-art MCMC algorithm and a workhorse for numerous high-dimensional practical inference problems (as mentioned in the references in \citep{Betancourt2017}). 

The main features/contributions/advantages of the proposed method are summarized below: 
\begin{enumerate}
    \item \textbf{Black-box simulators}: For many applications in science and engineering, the forward model is defined via a ``black-box'' simulation code. While these sophisticated black-box simulators deliver high-fidelity solutions, unfortunately, they are ill-suited when it comes to integrating with HMC. Here, the term ``black-box simulator'' refers to forward model simulators or computer programs ($\bm{f}$). These simulators take various problem parameters $\infv$ as input (such as geometric descriptions of the domain, boundary conditions, initial conditions, and material property distributions), and produce corresponding solutions $\meas = \bm{f}(\infv)$. However, their limitations become apparent in their inability to provide gradients of the simulator's output with respect to its inputs, i.e., $\partial \meas/\partial \infv$ remains unavailable. This constraint primarily arises from the complex and optimized nature of these simulators, which are often developed over years by multiple researchers, utilizing legacy programming languages (such as C or Fortran 77) and heavily optimized for forward computations, sometimes with parallel implementations \citep{harbaugh2005modflow, Keyes_etal_13, pflotran-paper}. These characteristics make it challenging to seamlessly integrate these simulators with modern automatic differentiation libraries or require substantial manual intervention to incorporate gradient functionality in them.  
    This bottleneck prevents the use of the standard HMC algorithm for such applications, as it requires reasonably accurate gradient information for robust performance. The method proposed in this manuscript, on the other hand, is perfectly suitable for such scenarios, as it can use such black-box simulator in conjunction with an appropriate differentiable surrogate model for efficient sampling.
    
    \item \textbf{Computation cost}: When applied to Bayesian inverse problems with PDE-based forward models, each step of a standard HMC algorithm requires $2\times L$ PDE solves (where, $L$=no.\ of leapfrog steps, and the factor of 2 accounts for one forward and one adjoint solve required for computing the gradient of the posterior). This can be computationally prohibitive for many practical problems. In contrast, the algorithm proposed in this manuscript can use a DNN-based surrogate model in the first stage and hence can leverage its automatic differentiation capabilities to accurately compute the required gradient at no additional cost. This approach is not only computationally cheaper but also much faster. Furthermore, the acceptance rate of the standard HMC algorithm is around 60--70\%~\citep{Beskos2013}, (i.e., 30 to 40\% of samples are discarded). This amounts to 2$\times$0.3$\times$L$\times$N to 2$\times$0.4$\times$L$\times$N ``wasted'' PDE solves (where $N$=no.\ of HMC samples). As demonstrated in the results section, our proposed algorithm significantly improves the acceptance rate, thus reducing the number of rejected samples and, as a result, significantly reducing the number of ``wasted'' PDE solves.
    
    \item \textbf{Statistical efficiency}: The proposed two-stage algorithm facilitates a longer Hamiltonian trajectory (without sacrificing acceptance rate) at each step due to the computationally inexpensive surrogate model in the first stage. This results in bigger jumps and a well-mixed Markov chain. Due to longer jumps, the effective sample size (ESS) of the resulting Markov chain is significantly larger. This means the samples are more uncorrelated, and hence it produces low-variance statistical estimates. This is something that is not practically feasible with the traditional HMC algorithm, as taking longer jumps often results in a significantly lower acceptance rate, which in turn increases the number of high-fidelity simulations required significantly, making it computationally prohibitive. Thus, for a given fixed compute budget of HF solves, with traditional HMC, one must choose between achieving a better ESS or a better acceptance rate. In contrast, MFHMC offers significantly (often orders of magnitude) better ESS \textit{and} acceptance rate simultaneously. This numerical superiority is demonstrated across a series of problems in the results section.

    \item $\textbf{Non-differentiable priors}$: In many applications the use of non-differentiable priors is widespread. This include image denoising~\citep{Rudin1992, Chambolle2010}, deblurring~\citep{Beck2009FastGA}, multi-frame superresolution~\citep{Farsiu1996FastAR}. Other than this, Laplace priors used in Bayesian Lasso regression~\citep{Park2012}, sample-based priors used in Bayesian inversion~\citep{Vauhkonen1997, patel2019bayesian}, Total-Variation and Bernoulli-Laplace priors used in sparse regularization~\citep{Chaari2013SparseBR, JJFern2004, lee2013bayesian} are popular examples of such priors. The lack of differentiability of resulting posterior distribution prevents using the HMC method for posterior exploration with such priors. This leads to two sub-optimal design choices: (i) selecting a non-gradient-based MCMC method for posterior exploration and/or (ii) selecting a second-choice prior, which is differentiable. In contrast, the method proposed in this manuscript does not face any such difficulties and can easily be deployed with non-differentiable priors.
\end{enumerate}

The remainder of this paper is organized as follows: first, we provide relevant background on the HMC algorithm and highlight some of the key features and challenges of this traditional HMC algorithm in Section \ref{sec:background}. Then we propose our novel MFHMC algorithm in Section \ref{sec:method_2hmc} and show how samples produced by it form a valid Markov Chain in Section \ref{sec:mfhmc_analysis}. In Section \ref{sec:results} we provide numerical results on a range of test problems involving synthetic and experimental data to demonstrate the proposed algorithm's effectiveness and versatility. Finally, we conclude the paper in Section \ref{sec:conclusion} with a brief summary of the paper with potential future directions.

\section{Background}\label{sec:background}
The method proposed in this manuscript is applicable to any sampling (from an unnormalized probability) task. However, to make ideas more concrete, we limit our focus on the sampling problem arising in Bayesian inverse problems. To this end, we first introduce the classical Bayesian inverse problem and relevant notations below and then explain how such problems can be solved using the traditional single-stage HMC algorithm, followed by our proposed two-stage MFHMC algorithm.

\subsection{Bayesian Inverse Problem and Markov Chain Monte Carlo}\label{sec:bayesian_inversion}
Consider the following direct/forward problem
\begin{equation}\label{eq:forward_model}
    \nmeas = \fhf(\infv) + \noise, \qquad \infv \in \Omega_{\mathcal{X}} \subset  \mathbb{R}^N, \qquad \nmeas \in \Omega_{\mathcal{Y}} \subset \mathbb{R}^M,
\end{equation}
where $\nmeas$ is the measured response to some input parameter $\infv$, $\fhf$ is the direct/forward model (often described via PDE), and $\noise$ represents measurement and/or modeling error. The inverse problem aims to recover the unknown parameter $\infv$ from the noisy and possibly partial measurement of $\nmeas$. Bayesian inference provides a principled probabilistic framework for solving such ill-posed problems with quantified uncertainty estimates. Within this approach, both the inferred field and the observation are modeled as a realization of random variables, $\mathcal{X}$ and $\mathcal{Y}$, respectively. Next, a prior probability density $\priorinfv(\infv)$, which captures all the constraints and the domain knowledge about the parameter to be inferred ($\mathcal{X}$) \textit{prior} to observing measurement ($\nmeas$) is assumed. Then this prior probability density is used in conjunction with the forward model ($\fhf$) to define the likelihood distribution of observing measurement $\mathcal{Y} = \nmeas$ given $\mathcal{X} = \infv$, that is $\plike(\nmeas|\infv)$. For additive measurement noise model, as described in \cref{eq:forward_model}, this likelihood distribution could be written as $\plike(\nmeas|\infv) = \pn(\nmeas - \fhf(\infv))$. Using Bayes' rule this yields the following expression for the posterior distribution of $\mathcal{X}$
\begin{align}
    \postinfv(\infv|\nmeas) = \frac{\plike(\nmeas|\infv)\priorinfv(\infv)}{\pmeas(\nmeas)} \propto \pn(\nmeas - \fhf(\infv))\priorinfv(\infv),
\end{align}
where $\pmeas(\nmeas)$ is called the evidence term, which normalizes the posterior distribution $\postinfv(\infv|\nmeas)$ so that $\int_\mathcal{X} \postinfv(\infv|\nmeas) \bm{d\infv} = 1$. Ideally, we would like to have access to the full posterior distribution as output to the Bayesian inference. However, since we lack the appropriate tools to visualize and understand this high-dimensional probability density, in practice, we mostly focus on computing lower-dimensional quantities of the posterior, such as the mean $\bar{\infv} = \mathbb{E}_{\infv\sim\postinfv}[\infv] = \int_\mathcal{X}\infv \postinfv(\infv|\nmeas)\bm{d\infv}$ or higher order moments, marginals, and confidence intervals.

For PDE-based inverse problems, however, this is problematic as for most of such problems the dimension of $\mathcal{X}$ is typically equal to the number of nodes (number of degrees of freedom to be more precise) in the numerical discretization scheme (such as finite element or finite volume method). For complex real-world problems, this could be as high as $\mathcal{O}(10^3)$--$\mathcal{O}(10^9)$ due to fine spatio-temporal discretization. Computing integral over such a high-dimensional space is simply infeasible using numerical integration (such as quadrature-based) techniques and the analytical treatment of this integral is not possible (unless for a simple and not-so-practical conjugate prior case). This motivates using MCMC methods that perform random walks in the parameters space and can generate samples according to the posterior probability distribution. Once these samples are generated, a simple Monte Carlo sum can be taken of these samples to approximate the integral.

\subsection{MCMC methods and Metropolis-Hastings}\label{sec:mcmc_mh}
Markov Chain Monte Carlo (MCMC) methods can generate samples from any unnormalized probability distribution, such as $\unnormpostinfv(\infv|\nmeas) := \pn(\nmeas - \fhf(\infv))\priorinfv(\infv)$. For this, MCMC methods construct a Markov Chain whose stationary distribution is the target unnormalized posterior density. This is achieved by assuming an initial distribution $\pi_0$ and a transition kernel $\mathcal{K}$, and constructing the following sequence of random variables:
\begin{equation}
    X_0 \sim \pi_0, \qquad X_{t+1} \sim \mathcal{K}(\cdot | X_t).
\end{equation}
In order for the $\unnormpostinfv$ to be the stationary distribution of the Markov Chain, three conditions must be satisfied: the kernel $\mathcal{K}$ must be irreducible and aperiodic (these are usually mild conditions and are satisfied) and $\unnormpostinfv$ has to be a fixed point of $\mathcal{K}$. This last condition can be expressed as: $p(\infv') = \int \mathcal{K}(\infv'|\infv)\unnormpostinfv(\infv)dx$. This condition is often satisfied by satisfying the detailed balance equation described as: $p(\infv')\mathcal{K}(\infv|\infv') = p(\infv)\mathcal{K}(\infv'|\infv)$.

Given any proposal distribution $q$, we can easily construct a transition kernel that respects detailed balance using Metropolis-Hastings accept/reject rules. More formally, starting from $\bm{x_0} \sim \pi_0$, at each step $t$, we sample $\infv'$ from a general transition probaility distribution $q(\cdot|X_t)$, and with probability 
$$
\alpha(\infv'|\bm{x_t}) = \text{min}\left(1, \frac{p(\infv')q(\bm{x_t}|\infv')}{p(\bm{x_t})q(\infv'|\bm{x_t})}\right),
$$
accept $\infv'$ as the next sample $\bm{x_{t+1}}$ in the chain and reject the sample $\infv'$ with the probability $1-\alpha(\infv'|\bm{x_t})$ and retain the previous state, i.e., $\bm{x_{t+1}}=\bm{x_t}$. For typical proposals, this algorithm converges to the target density in an asymptotic sense. However, this comes at the cost of very slow mixing (very slow traversal along the parameter space) as these algorithms are simply taking random jumps without considering the target distribution's geometry. Hamiltonian Monte Carlo (HMC) method tackles this problem of slow mixing by exploiting the geometry of the target density. Specifically, it uses the gradient information of the target density to make bigger proposal jumps in the parameter space along the regions of high probability.

\subsection{Hamiltonian Monte Carlo}\label{sec:method_1hmc}
In the Hamiltonian Monte Carlo algorithm (first introduced as ``Hybrid Monte Carlo'' in \citet{duane1987hybrid}), the state space to be inferred ($\infv$) is augmented with a fictitious momentum variable $\mome \in \mathbb{R}^N$. Then the canonical joint posterior density is defined as
\begin{align*}
    p^{\mathrm{post}}(\infv, \mome|\nmeas) \propto \mathrm{exp}\{-H(\infv, \mome)\}, 
\end{align*}
where 
$$H(\infv, \mome) := -\log\, p^{\mathrm{post}}(\infv|\nmeas) + \frac{\|\mome\|^2}{2}$$
is the Hamiltonian of the system. Here, the Hamiltonian $H(\infv, \mome) := U(\infv) + K(\mome)$, where the potential energy $U(\infv) := -\log\, p^{\mathrm{post}}(\infv|\nmeas)$ is a function of $\infv$ only and the kinetic energy $K(\mome) := \frac{\|\mome\|^2}{2}$ is a function of $\mome$ only. This separation of variables helps in defining the update rule for each variable. Interested readers are referred to \citet{Neal2012} for a more detailed discussion on HMC.
\begin{algorithm}[htbp]
\SetAlgoLined
\KwIn{$U(\bm{x}) = -\log p^{\mathrm post}(\infv|\nmeas), K(\bm{\xi}) = \bm{\xi}^T\bm{\xi}/2$, step size ($\epsilon$), number of leapfrog steps ($L$), number of HMC steps ($m$)}
\KwResult{$\bm{x_1}, \bm{x_2}, \cdots, \bm{x_m}$}
 Choose a starting point $\bm{x_1}$
 
 \For{i = 1, $\cdots$, m-1}{
    Set $\bm{x_i^{(0)}} = \bm{x_i}$
    
    Draw $\bm{\xi_0} \sim \mathcal{N}(\bm{0}, \bm{\mathds{1}})$
    
    Make half-step update of $\bm{\xi}$
    
    $\qquad \bm{\xi^{(0)}} = \bm{\xi_0} - \epsilon \textcolor{magenta}{\nabla_{\bm{x}} U(\bm{x_i^{(0)}})}/2 $
    
    \For{l = 1, $\cdots$, L-1}{
    Make full-step update of $\bm{x}$
    
    $\bm{x_i^{(l)}} = \bm{x_i^{(l-1)}} + \epsilon \bm{\xi^{(l-1)}}$
    
    Make full-step update of $\bm{\xi}$
    
    $\bm{\xi^{(l)}} = \bm{\xi^{(l-1)}} - \epsilon \textcolor{magenta}{\nabla_{\bm{x}} U(\bm{x_i^{(l)}})}$ 
    }

    Make final update of $\bm{x}$
    
    $\qquad \bm{x_i^{(L)}} = \bm{x_i^{(L-1)}} + \epsilon \bm{\xi^{(L-1)}}$
    
    Make final half-step update of $\bm{\xi}$
    
    $\qquad \bm{\xi^{(L)}} = \bm{\xi^{(L-1)}} - \epsilon \textcolor{magenta}{\nabla_{\bm{x}} U(\bm{x_i^{(L)}})}/2 $
    
    Negate momentum at the end of trajectory to make the proposal symmetric $\bm{\xi^{(L)}} = - \bm{\xi^{(L)}}$
    
    Compute acceptance probability
    $\qquad \alpha(\bm{x_i^{(0)}}, \bm{x_i^{(L)}}) = \mathrm{min} \left\{1, \exp(-U(\bm{x_i^{(L)}}) - K(\bm{\xi^{(L)}}) + U(\bm{x_i^{(0)}}) + K(\bm{\xi^{(0)}})) \right\}$
    
    Set $\bm{x_{i+1}}$ to

    $\qquad \bm{x_{i+1}}= 
    \begin{cases}
        \bm{x_i^{(L)}},      & \text{with probability } \alpha(\bm{x_i^{(0)}}, \bm{x_i^{(L)}})\\
        \bm{x_i^{(0)}},    & \text{with probability } 1-\alpha(\bm{x_i^{(0)}}, \bm{x_i^{(L)}})
    \end{cases}$
 }
 Return $\bm{x_1}, \bm{x_2}, \cdots, \bm{x_m}$
 
 \caption{Single Stage Hamiltonian Monte Carlo Algorithm}
 \label{alg:hmc}
\end{algorithm}

The detailed algorithm for HMC is provided in Alg. \ref{alg:hmc}. As highlighted in this algorithm (in magenta) a single step of HMC requires multiple evaluations of  $\nabla_{\infv}U(\infv)$. This requires having access to the gradient of the posterior distribution since $\nabla_{\infv}U(\infv) = -\nabla_{\infv} \log \postinfv(\infv|\nmeas)$. This, in turn, requires expensive computation of the gradient of the forward model since $ -\nabla_{\infv} \log \postinfv(\infv|\nmeas) = -\nabla_{\infv}\log \priorinfv(\infv) -\nabla_{\infv} \log \plike(\nmeas - \fhf(\infv))$. This is not tractable for black-box/expensive forward models. The two-stage Multi-fidelity HMC algorithm proposed below tackles this problem. 

    
    
    

\section{Multi-fidelity HMC (MFHMC)}\label{sec:method_2hmc}
Let $\flf$ denote the surrogate forward model for $\fhf$, and let $\postlf(\infv|\nmeas) \propto \pn(\nmeas - \flf(\infv))\priorinfv(\infv)$ and $\posthf(\infv|\nmeas) \propto \pn(\nmeas - \fhf(\infv))\priorinfv(\infv)$ denote the posterior density induced by the surrogate forward model and the high-fidelity forward model, respectively. Typical examples of $\fhf$ may include forward models defined by accurate numerical solvers (based on finite element or finite volume methods, for example), while $\flf$ may include computationally inexpensive and relatively less accurate surrogate models (based on proper orthogonal decomposition or deep neural networks, for example).

The overall idea of the proposed two-stage MFHMC algorithm as outlined in Section \ref{sec:intro} and \cref{fig:workflow} is as follows:
\begin{enumerate}[\textit{Stage} 1.]
    \item We use $\flf$ (and corresponding $\postlf(\infv|\nmeas)$) in a standard HMC iteration (described above) as the first stage. So, at $i^{th}$ iteration/step of MFHMC (with $\infv_{i-1}$ as current inferred variable state and $\mome_0$ as momentum variable), we propose $(\infv^{LF}, \mome)$ by following the Hamiltonian dynamics steps.    
    And we set $\bm{x^{HF}}$ to
    \begin{align}\label{eq:xhf}
        \bm{x^{HF}}= 
        \begin{cases}
            \bm{x^{LF}},      & \text{with probability } \alpha^{LF}( \bm{x_{i-1}}, \bm{x^{LF}})\\
            \bm{x_{i-1}},    & \text{with probability } 1-\alpha^{LF}( \bm{x_{i-1}}, \bm{x^{LF}})
        \end{cases}
    \end{align}
    where, $\alpha^{LF}(\bm{x_{i-1}}, \bm{x^{LF}})$ is the acceptance probability for this first (LF) stage and is defined as
    \begin{align}\label{eq:alpha_lf}
        \alpha^{LF}(\bm{x_{i-1}}, \bm{x^{LF}}) = \mathrm{min} \left\{1, \exp(-\ulf(\bm{x^{LF}}) - K(\bm{\xi}) + \ulf(\bm{x_{i-1}}) + K(\bm{\xi_{0}})) \right\}
    \end{align}
    
    Note that this is the standard acceptance probability for HMC (as also shown in Section \ref{sec:method_1hmc}). The only difference being the potential energy is computed using $\flf$, i.e., $\ulf(\infv):=-\log\postlf(\infv)$.
    \item Next, we accept $\bm{x^{HF}}$ as a sample with probability
    \begin{align}\label{eq:alpha_hf}
        \alpha^{HF}(\bm{x_{i-1}}, \bm{x^{HF}}) = \mathrm{min} \left\{1, \frac{p^{\mathrm post(HF)}(\bm{x^{HF}}|\nmeas) \postlf(\bm{x_{i-1}}|\nmeas)}{p^{\mathrm post(HF)}(\bm{x_{i-1}}|\nmeas)\postlf(\bm{x^{HF}}|\nmeas)} \right\}
    \end{align}
    i.e., set $\bm{x_{i}} = \bm{x^{HF}}$ with probability $ \alpha^{HF}(\bm{x_{i-1}}, \bm{x^{HF}})$ and $\bm{x_{i}} = \bm{x_{i-1}}$ with probability $ 1-\alpha^{HF}(\bm{x_{i-1}}, \bm{x^{HF}})$.    
\end{enumerate}

\begin{algorithm}[htbp]
\SetAlgoLined
\KwIn{$p^{\mathrm post (HF)}(\infv|\nmeas)$, $p^{\mathrm post (LF)}(\infv|\nmeas)$, step size ($\epsilon$), number of leapfrog steps ($L$), number of HMC steps ($m$), $\ulf(\infv) := -\log p^{\mathrm post (LF)}$, $K(\bm{\xi}) := \bm{\xi}^T\bm{\xi}/2$}
\KwResult{$\bm{x_1}, \bm{x_2}, \cdots, \bm{x_m}$}
 Choose a starting point $\bm{x_0}$
 
 \For{i = 1, $\cdots$, m}{
    \begin{mdframed}[backgroundcolor=lightcyan!50, roundcorner=1pt, linewidth=0pt]
     \hspace{\stretch{1}} \textbf{\underline{First Stage}} \vspace{-\baselineskip} 
 
    $\bm{x^{LF}} = \bm{x_{i-1}}$ 
    
    Draw $\bm{\xi_0} \sim \mathcal{N}(\bm{0}, \bm{\mathds{1}})$
    
    Make half-step update of $\bm{\xi}$    
    
    $\qquad \bm{\xi} = \bm{\xi_0} - \epsilon \nabla_{\bm{x}} \ulf(\bm{x^{LF}})/2 $
      
    \For{l = 1, $\cdots$, L-1}{
        Make full step update of $\infv$        
        
        $\bm{x^{LF}} = \bm{x^{LF}} + \epsilon \bm{\xi}$        
        
        Make full step update of $\bm{\xi}$        
        
        $\bm{\xi} = \bm{\xi} - \epsilon \nabla_{\bm{x}} \ulf(\bm{x^{LF}})$ 
    }
    Make last half-step update of $\infv$
    
    $\qquad \bm{x^{LF}} = \bm{x^{LF}} + \epsilon \bm{\xi}$
    
    Make half-step update of $\bm{\xi}$
    
    $\qquad \bm{\xi} = \bm{\xi} - \epsilon \nabla_{\bm{x}} \ulf(\bm{x^{LF}})/2 $
    
    Negate momentum at the end of trajectory to make the proposal symmetric 
    
    $\qquad \bm{\xi} = - \bm{\xi}$
   
    Compute acceptance probability for the first (LF) stage
    
    $ \alpha^{LF}(\bm{x_{i-1}}, \bm{x^{LF}}) = \mathrm{min} \left\{1, \exp(-\ulf(\bm{x^{LF}}) - K(\bm{\xi}) + \ulf(\bm{x_{i-1}}) + K(\bm{\xi_{0}})) \right\}$
      
    Set $\bm{x^{HF}}$ to
    
    $\qquad \bm{x^{HF}}= 
    \begin{cases}
        \bm{x^{LF}},      & \text{with probability } \alpha^{LF}( \bm{x_{i-1}}, \bm{x^{LF}})\\
        \bm{x_{i-1}},    & \text{with probability } 1-\alpha^{LF}( \bm{x_{i-1}}, \bm{x^{LF}})
    \end{cases}$
    \end{mdframed}
    \vspace{-1.2\baselineskip}
    \begin{mdframed}[backgroundcolor=lightmauve!50, roundcorner=1pt, linewidth=0pt]
    \hspace{\stretch{1}} \textbf{\underline{Second Stage}} \vspace{-\baselineskip} 
     
    Compute acceptance probability for the second (HF) model 
    
    $\qquad \alpha^{HF}(\bm{x_{i-1}}, \bm{x^{HF}}) = \mathrm{min} \left\{1, \frac{p^{\mathrm post(HF)}(\bm{x^{HF}}|\nmeas) \postlf(\bm{x_{i-1}}|\nmeas)}{p^{\mathrm post(HF)}(\bm{x_{i-1}}|\nmeas)\postlf(\bm{x^{HF}}|\nmeas)} \right\}$
    
      
    Set sample $\bm{x_{i}}$ to
    
    $\qquad \bm{x_i}= 
    \begin{cases}
    \bm{x^{HF}},      & \text{with probability } \alpha^{HF}(\bm{x_{i-1}}, \bm{x^{HF}}) \\
    \bm{x_{i-1}},    & \text{with probability } 1-\alpha^{HF}(\bm{x_{i-1}}, \bm{x^{HF}})
    \end{cases}$
    \end{mdframed}
 }
 Return $\bm{x_1}$, $\bm{x_2}$, $\cdots$, $\bm{x_m}$ 
 \caption{Multi-fidelity Hamiltonian Monte Carlo (MFHMC) Algorithm}
 \label{alg:mfhmc}
\end{algorithm}
Note that in the above two-stage algorithm, if the proposed sample at the end of Hamiltonian trajectory $\bm{x^{LF}}$ is rejected by the first stage, $\bm{x_{i-1}}$ will be sent to the second stage (i.e., $\bm{x^{HF}} = \bm{x_{i-1}}$). Now, since $\alpha^{HF}(\bm{x_{i-1}}, \bm{x_{i-1}}) = 1$, no further high-fidelity computation is needed in the second stage and we can simply set $\bm{x_i}=\bm{x_{i-1}}$. Thus, the expensive high-fidelity solve can be avoided for the samples which are unlikely to be accepted. In contrast, traditional HMC (Section \ref{sec:method_1hmc}) require $2\times L$ high-fidelity evaluation for every proposed state even if that proposed state is rejected in the end. 

Moreover, in this two-stage algorithm, gradient computation is solely required in the first stage (for the HMC proposal step), where the surrogate model $\flf$ is used. This surrogate is selected in such a way that its gradient can be computed in an expensive yet fast manner (using automatic differentiation, for example), whereas in the second stage (Metropolis-Hastings step) where $\fhf$ is used, no gradient calculation is required. Hence, accurate black-box forward model simulators can also be used for this stage. One can think of the first stage of the MFHMC algorithm as a first-pass filter that filters out ``bad'' samples, which are less likely to be accepted and thus saves expensive high-fidelity solves for them, whereas the second stage corrects the error introduced by the use of the surrogate model in the first stage for posterior calculation and results in accurate sampling from the true posterior. This is ensured by the acceptance probability for the second stage (\cref{eq:alpha_hf}). We show in the next section that this acceptance probability ($\alpha^{HF}(\cdot, \cdot)$) satisfies the detailed equation (\cref{eq:detailed_balance_mfhmc}) and results in a valid Markov chain. The detailed two-stage MFHMC algorithm is provided in Alg. \ref{alg:mfhmc}.

\subsection{Analysis of MFHMC}\label{sec:mfhmc_analysis}
Next, we will analyze the MFHMC algorithm in more detail following~\citep{Efendiev2006}. Denote
\begin{align*}
    \mathcal{E} &= \{(\infv, \mome); \posthf(\infv, \mome) > 0 \}, \\
    \mathcal{E}^* &= \{(\infv, \mome); \postlf(\infv, \mome) > 0 \}, \\
    \mathcal{D} &= \{(\infv, \mome); q(\infv, \mome | \infv_{i-1}, \mome_{i-1}) > 0 \text{ for some } (\infv_{i-1}, \mome_{i-1}) \in \mathcal{E}\}.
\end{align*}

The set $\mathcal{E}$ is the support of the target posterior distribution $\posthf(\infv, \mome)$. It contains all the possible tuples of the inferred variable and momentum variable $(\infv, \mome)$ which has a positive probability of being accepted as a sample. Similarly, $\mathcal{E}^*$ is the support of the low fidelity posterior distribution  $\postlf(\infv, \mome)$, which contains all tuples $(\infv, \mome)$ which can be accepted by the first stage of our algorithm. $\mathcal{D}$ contains proposals which can be generated by the proposal distribution $q(\infv, \bm{\xi} | \infv_{i-1}, \bm{\xi_{i-1}})$. For the MFHMC algorithm to work properly, the following two conditions must hold all the time: $\mathcal{E} \subseteq \mathcal{D}$ and $\mathcal{E} \subseteq \mathcal{E}^*$. If one of these conditions is not true, for example, say, $\mathcal{E} \not\subseteq \mathcal{E}^*$, then there will exist a subset $K \subset (\mathcal{E} \backslash \mathcal{E}^*)$ such that
\[
    \posthf(K) = \int_K \posthf(\infv, \mome) \bm{dxd\xi} > 0 \: \text{ and } \: \postlf(K) = \int_K \postlf(\infv, \mome) \bm{dxd\xi} = 0
\]
which means no elements of $K$ can pass the first stage of MFHMC and $K$ will never be visited by the final chain despite its samples having a positive probability of being accepted by $\posthf$ and hence the resulting Markov chain will not be sampled properly. 

For most practical purposes, the conditions $\mathcal{E}, \mathcal{E}^* \subset \mathcal{D}$ can be naturally satisfied by selection of appropriate proposal distribution $q(\infv, \mome | \infv_{i-1}, \mome_{i-1})$. By choosing the appropriate value of the likelihood variance in $\postlf$ the condition $\mathcal{E} \subset \mathcal{E}^*$ can also be satisfied. Thus $\mathcal{E} \subset \mathcal{E}^* \subset \mathcal{D}$. In this case, $\mathcal{E}^*$ is identical to the support of the transition probability distribution $Q(\infv, \mome | \infv_{i-1}, \mome_{i-1}) = \alpha^{\text{(LF)}}((\infv_{i-1}, \mome_{i-1}), (\infv, \mome))q(\infv, \mome | \infv_{i-1}, \mome_{i-1})$:
\begin{align*}
    \mathcal{E}^* = \{(\infv, \mome); Q(\infv, \mome | \infv_{i-1}, \mome_{i-1}) > 0 \text{ for some } (\infv_{i-1}, \mome_{i-1}) \in \mathcal{E}\}.
\end{align*}

Due to the very high dimension of the joint field $(\infv, \mome)$, the support $\mathcal{E}$ of the target distribution $\posthf$ is much smaller than the support $\mathcal{D}$ of the proposal distribution $q(\infv, \mome | \infv_{i-1}, \mome_{i-1})$. For all the proposed samples $(\infv, \mome) \in (\mathcal{D} \backslash \mathcal{E})$, they will never be accepted as valid samples in the final Markov chain in the traditional single-stage HMC algorithm resulting in huge computation waste since $\posthf(\infv, \mome) = 0$. In the proposed MFHMC algorithm, however, the transition probability distribution $Q(\infv, \mome | \infv_{i-1}, \mome_{i-1})$ (which acts as an effective proposal distribution for the second stage) samples from a much smaller support $\mathcal{E}^*$, hence avoids solving expensive HF problem for all $(\infv, \mome) \in \mathcal{D}\backslash\mathcal{E}^*$. For each iteration, the MFHMC algorithm only requires the HF simulation $r$ times in average, where 
\begin{align*}
    r = \int_{\mathcal{E}^*} \alpha^{\text{(LF)}}((\infv_{i-1}, \mome_{i-1}), (\infv, \mome))q(\infv, \mome | \infv_{i-1}, \mome_{i-1}) < 1.
\end{align*}

Note that $\int_{\mathcal{D}} q(\infv, \mome | \infv_{i-1}, \mome_{i-1})=1$ and $\alpha^{\text{(LF)}}((\infv_{i-1}, \mome_{i-1}), (\infv, \mome)) \leq 1$. If $\mathcal{E}^*$ is close to $\mathcal{E}$ and hence much smaller than $\mathcal{D}$, then $r \ll 1$. Therefore, the proposed MFHMC method requires much fewer HF simulations while approximately still accepting the same amount of proposals. In other words, the MFHMC algorithm can achieve a much higher acceptance rate for each HF simulation (as also demonstrated via numerical experiments in Section \ref{sec:results}).

\subsection*{Detailed balance for MFHMC} Next we will discuss the stability properties of the MFHMC algorithm and show that it shares the same convergence property as the traditional HMC algorithm. For this, we will show that the resulting Markov chain is ergodic, irreducible, and aperiodic by proving the satisfaction of the detailed balance equation. Denote by $\mathcal{K}$ the transition kernel of the Markov chain $\{\bm{x_i}\}$ generated by the MFHMC algorithm and let augmented variable $\joint:=(\infv, \mome)$ denote the current state of both inferred variable $\infv$ and momentum variable $\mome$. Since the effective proposal distribution is given by
\begin{align}
    Q(\joint | \joint_{i-1}) = \alpha^{\text{(LF)}}(\joint_{i-1}, \joint)q(\joint | \joint_{i-1}) +
    \left(1 - \int \alpha^{\text{(LF)}}(\joint_{i-1}, \joint)q(\joint | \joint_{i-1})\right)\delta_{\joint_{i-1}}(\joint)
\end{align}

The transition kernel of the overall Markov chain is given by
\begin{align}
    \mathcal{K}(\joint_{i-1}, \joint) &= \alpha^{\text{(HF)}}(\joint_{i-1}, \joint)Q(\joint | \joint_{i-1}) \quad \text{ for } \joint \neq \joint_{i-1} \label{eq:transition_kernel_mfhmc}\\
    \mathcal{K}(\joint_{i-1}, \{\joint\}) &= 1 - \int_{\joint \neq \joint_{i-1}} \alpha^{\text{(HF)}}(\joint_{i-1}, \joint)Q(\joint | \joint_{i-1}) \quad \text{ for } \joint = \joint_{i-1} 
\end{align}
It is easy to show that this transition kernel satisfies the detailed balance equation
\begin{align}\label{eq:detailed_balance_mfhmc}
    \posthf(\joint_{i-1}) \mathcal{K}(\joint_{i-1}, \joint) = \posthf(\joint) \mathcal{K}(\joint, \joint_{i-1}) 
\end{align}
for any $\joint, \joint_{i-1} \in \mathcal{E}$. The above equality is obviously true when $\joint = \joint_{i-1}$. When $\joint \neq \joint_{i-1}$ then from the definition of the transition kernel of the overall Markov chain \cref{eq:transition_kernel_mfhmc} we have
\begin{align*}
    \posthf(\joint_{i-1}) \mathcal{K}(\joint_{i-1}, \joint) &= \posthf(\joint_{i-1}) \alpha^{\text{(HF)}}(\joint_{i-1}, \joint)Q(\joint | \joint_{i-1}) \\
    &= \posthf(\joint_{i-1})Q(\joint | \joint_{i-1})\mathrm{min}\left\{1, \frac{\posthf(\joint)\postlf(\joint_{i-1})}{\posthf(\joint_{i-1})\postlf(\joint)}\right\} \\
    &= \mathrm{min}\left\{Q(\joint|\joint_{i-1})\posthf(\joint_{i-1}), Q(\joint_{i-1}|\joint)\posthf(\joint)) \right\} \\
    &= \mathrm{min}\left\{\frac{Q(\joint|\joint_{i-1})\posthf(\joint_{i-1})}{Q(\joint_{i-1}|\joint)\posthf(\joint))}, 1 \right\} Q(\joint_{i-1}|\joint)\posthf(\joint)) \\
    &= \alpha^{\mathrm{(HF)}}(\joint, \joint_{i-1}) Q(\joint_{i-1}|\joint)\posthf(\joint)) \\
    &= \posthf(\joint) \mathcal{K}(\joint, \joint_{i-1}).
\end{align*}
So the detailed balance is always satisfied. Using the above relation, we can easily show that 
$$\posthf(A) = \int \mathcal{K}(\joint, A) \bm{ds}$$
for any $A \in \mathcal{B}(A)$, where $\mathcal{B}$ denotes all the Borel measurable subset of $\mathcal{E}$ and hence $\posthf$ is indeed the stationary distribution of $\mathcal{K}$.

\section{Numerical Validation and Results}\label{sec:results}
In this section, we present the numerical results obtained through the application of the proposed MFHMC algorithm. We initiate this exploration with a systematic study focused on an analytically tractable high-dimensional target distribution. Through careful selection of various low-fidelity models, we conduct performance comparisons between MFHMC and HMC across different computational budgets, as determined by the number of target density evaluations. Subsequently, we consider both linear as well as non-linear Bayesian inverse problems commonly encountered in computational science and engineering.

\textbf{Inference problems and datasets}: We investigate four distinct inference problems to showcase the versatility and applicability of the MFHMC algorithm across different scenarios: (i) Sampling from a 250-dimensional ill-conditioned multi-variate normal distribution (Section \ref{sec:250d}), (ii) Initial condition inversion problem for the transient diffusion equation (Section \ref{sec:init_cond_inv}), (iii) Coefficient inversion problem for an elliptical PDE (Section \ref{sec:coefficient_inv}), (iv) Hydraulic tomography (a coefficient inversion problem in Darcy's flow with mixed boundary condition (Section \ref{sec:real_tomography})). Our experimentation embraces diverse high- and low-fidelity forward models, diverse prior models, and real-world datasets, ensuring a comprehensive assessment:
\begin{enumerate}
    \item \textit{High-fidelity (HF) forward models} encompass finite difference-based (Section \ref{sec:init_cond_inv}) and finite element-based (Section \ref{sec:coefficient_inv}, \ref{sec:real_tomography}) solvers, mirroring common practices in scientific and engineering domains.
    \item \textit{Low-fidelity (LF) forward models} include truncated SVD-based (Section \ref{sec:init_cond_inv_gaussian_prior}, \ref{sec:init_cond_inv_gan_prior}) and deep learning-based surrogate models leveraging fully connected (Section \ref{sec:init_cond_inv_gan_prior}) and convolutional networks (Section \ref{sec:coefficient_inv}, \ref{sec:real_tomography}).
    \item \textit{Different prior models}  such as Gaussian priors (Section \ref{sec:init_cond_inv_gaussian_prior}) and data-driven/deep generative priors (Section \ref{sec:init_cond_inv_gan_prior}, \ref{sec:coefficient_inv}, \ref{sec:real_tomography}) offer insights into different Bayesian modeling strategies.
    \item \textit{Diverse Datasets} for the inferred and measured field for different problems including parametric (rectangular) dataset with simulated measurement (Section \ref{sec:init_cond_inv}, \ref{sec:darcy_rectangle}), channelized flow dataset mimicking subsurface groundwater channel with simulated measurement (Section \ref{sec:permeability_inversion}), and the experimental dataset obtained from a lab-scale study (Section \ref{sec:real_tomography})  ensure robust validation across different scenarios.
\end{enumerate}
This methodical diversity substantiates the algorithm's adaptability and extends the relevance of our findings to broad practical contexts.

\textbf{Baseline}: For baseline we consider a single-stage HMC algorithm, which is currently one of the state-of-the-art MCMC algorithms. For both HMC and MFHMC we use burn-in period of 0.25.

\textbf{Performance evaluation metrics}: In order to compare the relative performance of the proposed MFHMC algorithm to the baseline standard single-stage HMC algorithm, we consider four different metrics for computational efficiency and accuracy. These are: the number of accepted moves per HF evaluations (Accepted moves/$n_{hf}$), effective sample size per HF evaluations (ESS/$n_{hf}$), the relative error in posterior statistics, and the ESJD (expected square jump distance) per HF evaluations (ESJD/$n_{hf}$). A detailed description of these metrics is provided below.

\begin{table}
  \centering
  \begin{tabular}{|p{4.5cm}|p{4.25cm}|p{7.05cm}|}
    \toprule
    \textbf{Metric} & \textbf{Formula} & \textbf{Descriptive details} \\
    \midrule
    No. of accepted moves & \multirow{2}{*}{Accepted moves/$n_{hf}$} & Accepted moves = \# of accepted HF \\
    per HF evaluation & & moves;  $n_{hf}$ = \# of HF evaluations \\ 
    \midrule[\lightrulewidth] 
    Effective sample size & \multirow{2}{*}{ESS/$n_{hf}$} & \multirow{2}{*}{Eq.\eqref{eq:ess}} \\ 
    per HF evaluation & & \\ 
    \midrule[\lightrulewidth] 
    Relative Error in & \multirow{2}{*}{$\frac{||\bm{v_{MCMC}} - \bm{v_{true}}||_2}{||\bm{v_{true}}||_2}\times 100 \%$} & $\bm{v}$ is a posterior statistic \\
    posterior stat. (in \%) & & such as mean or covariance \\ 
    \midrule[\lightrulewidth] 
    Expected Squared Jump & \multirow{2}{*}{ESJD/$n_{hf}$} & \multirow{2}{*}{Eq.\eqref{eq:esjd}} \\
    Distance per HF eval. & & \\
    \bottomrule
  \end{tabular}
  \caption{Performance evaluation metrics}
\end{table}
\textbf{Effective Sample Size}:
\begin{align}\label{eq:ess}
    \text{ESS}/n_{hf} = \min_{d=\{1, \cdots, D\}}\frac{M}{n_{hf}\left(1 + 2 \sum_{s=1}^{M-1}(1-\frac{s}{M})\rho^d_s\right)}
\end{align}
where,
\begin{align*}
    \rho^d_s = \text{Cov}(x^d_t, x^d_{t+s})/\text{Var}(x^d_t)
\end{align*}
Here, $\bm{x_1}, \bm{x_2}, \cdots, \bm{x_M}$ refers to the post burn-in samples of an MCMC chain of length $M$ and dimension $D$, and $n_{hf}$ refers to the number of HF evaluations. $\text{Cov}(\cdot)$ and $\text{Var}(\cdot)$ refers to covariance and variance. 

Note that following~\citep{parno2018transport}, while the autocorrelation calculation ($\rho^d_s$) in \cref{eq:ess} uses only the samples produced after the burn-in period, the normalized values of ESS reported in all numerical experiments (ESS/$n_{hf}$) use \textit{all} HF evaluations (for $n_{hf}$ value). Thus, the cost of burn-in is reflected in these normalized performance metric. Additionally, the minimum value of ESS is reported across the dimensions $D$ of the Markov chain as a conservative estimate of the overall effective sample size, as this dimension typically determines the overall convergence and reliability of the Markov chain.
 
\textbf{Expected Squared Jump distance}:
\begin{align}\label{eq:esjd}
    \text{ESJD}/n_{hf} = \frac{\mathbb{E}_{\pi}\left[\lVert \bm{x_{t+1}} - \bm{x_t} \rVert^2\right]}{n_{hf}}
\end{align}
where $\pi$ refers to the distribution or measure, which represents the underlying probability distribution of the Markov Chain.

\subsection{250-dimensional ill-conditioned multivariate normal (MVN)}\label{sec:250d}
This study aims to systematically quantitatively evaluate MFHMC versus HMC across various fixed computation budgets. To achieve this, we focus on a problem where the target distribution (along with its moments such as mean and covariance) is known. This knowledge enables the computation and comparison of the accuracy of these moments.

Specifically, we consider a 250-dimensional ill-conditioned multivariate Gaussian distribution with zero mean and a known precision matrix $A^{HF}$, i.e.,
\begin{equation}\label{eq:mult_gaussian_hf}
    \posthf(\infv) := \mathcal{N}(\bm{0}, \Sigma^{HF}={A^{HF}}^{-1}) \propto \exp\left(-\frac{1}{2}\infv^T A^{HF} \infv\right).
\end{equation}
The matrix $A^{HF}$ was generated from Wishart distribution with an identity scale matrix and 250 degrees of freedom. This yields a target distribution with a strongly correlated covariance matrix ($\Sigma^{HF}={A^{HF}}^{-1}$). A similar target distribution is used in previous studies \citep{hoffman2014no, hoffman2021adaptive}.

We consider multiple LF target distributions to assess the effect of the fidelity of the LF model on the overall performance of MFHMC. These different LF distributions are parameterized by a scalar parameter $\gamma$ as follows:
\begin{equation*}
    \Sigma^{LF} := \Sigma^{HF} + \frac{\gamma}{d} \times \text{trace}(\Sigma^{HF}) \times I(d).
\end{equation*}
where $d=250$ is the dimension of $\infv$ and $I(d)$ denotes identity matrix of dimension 250. Thus, by changing the value of $\gamma$, we get different $\Sigma^{LF}$ and as a consequence different 
$$\postlf(\infv) := \mathcal{N}(\bm{0}, \Sigma^{LF}={A^{LF}}^{-1}).$$
We consider four different LF models by selecting $\gamma$ = [$10^{-4}, 10^{-5}, 10^{-6}, 10^{-7}$] which corresponds to [$17.65\%, 2.14\%, 0.22\%, 0.02\%$] relative error in precision matrix respectively (defined as $\left(\frac{A^{LF} - A^{HF}}{A^{HF}}\right)\times100$). 

\begin{figure}[htbp]
    \centering
    \includegraphics[width=0.8\linewidth]{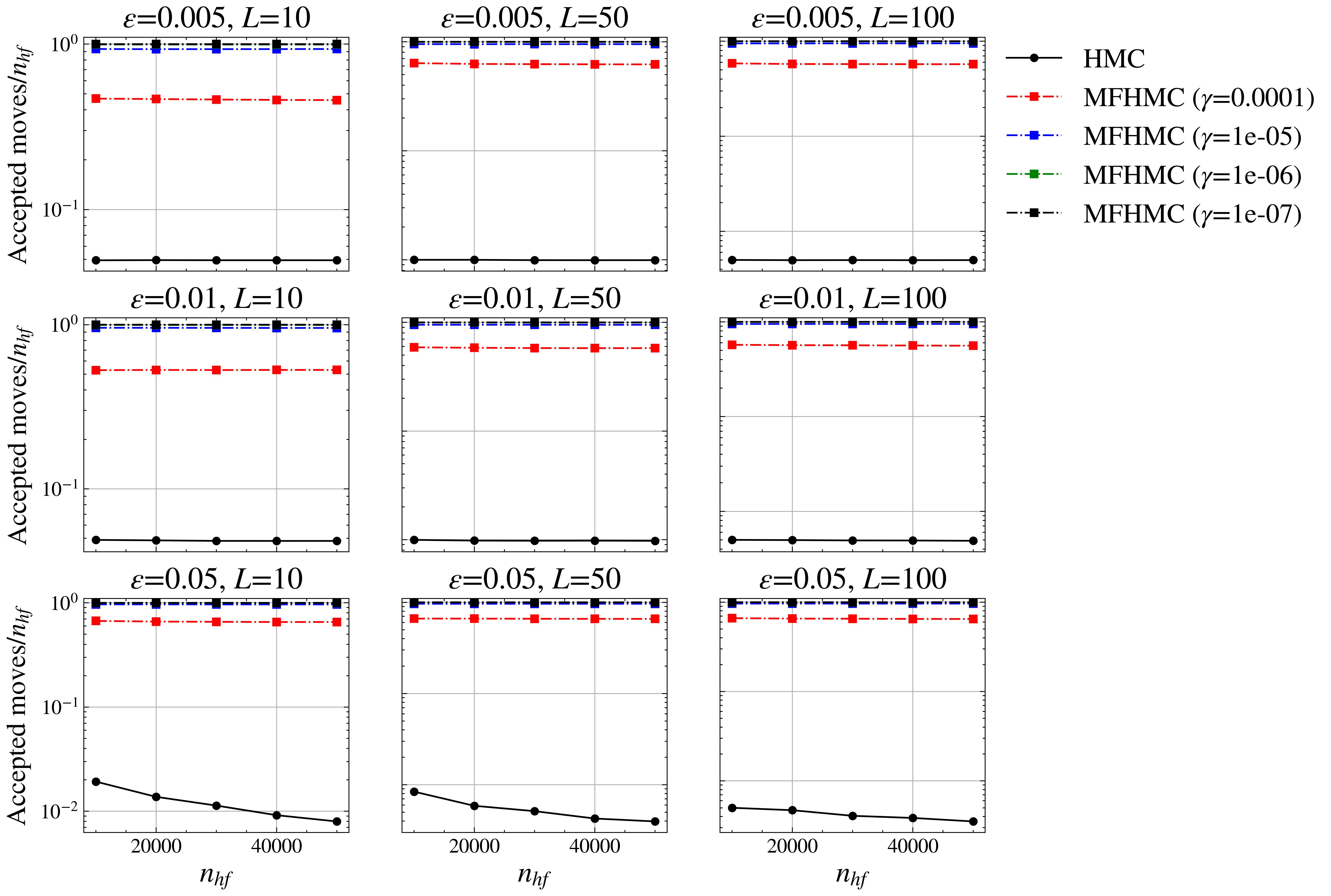}
    \caption{Comparison of the number of accepted moves per HF evaluation for HMC and MFHMC at different computation budgets (as defined by the number of HF target distribution evaluations ($n_{hf}$)). To minimize the influence of randomness, each experiment was performed with five different random seeds. Each data point in the figure represents the average value of these five experiments.}
    \label{fig:ar_vs_hf_250mvn}
\end{figure}

\begin{figure}[htbp]
    \centering
    \includegraphics[width=0.8\linewidth]{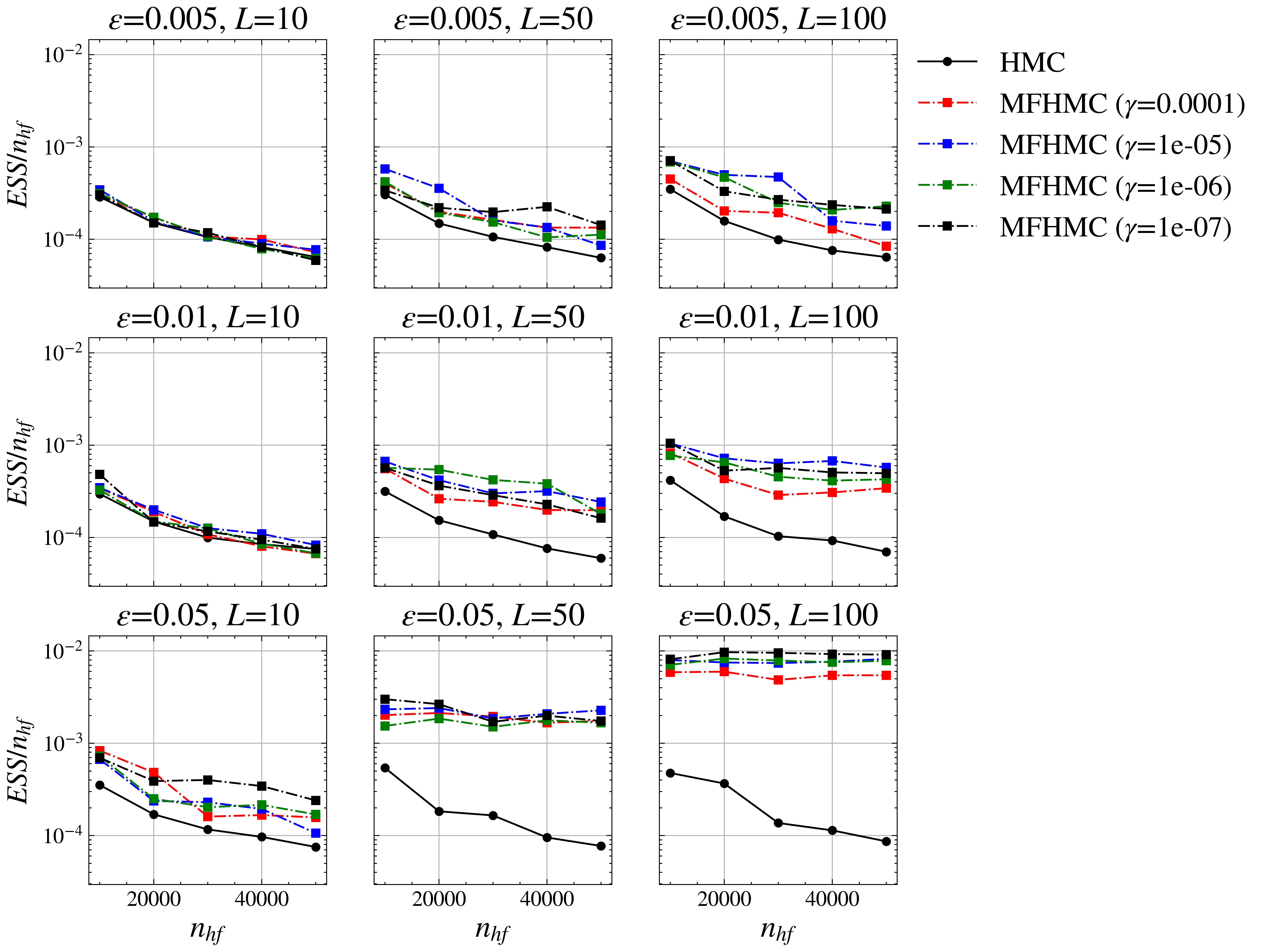}
    \caption{Comparison of effective sample size (ESS) per HF evaluation for HMC and MFHMC at different computation budgets (as defined by the number of HF target distribution evaluations ($n_{hf}$)). To minimize the influence of randomness, each experiment was performed with five different random seeds. Each data point in the figure represents the average value of these five experiments.}
    \label{fig:ess_vs_hf_250mvn}
\end{figure}


\begin{figure}[htbp]
    \centering
    \includegraphics[width=0.8\linewidth]{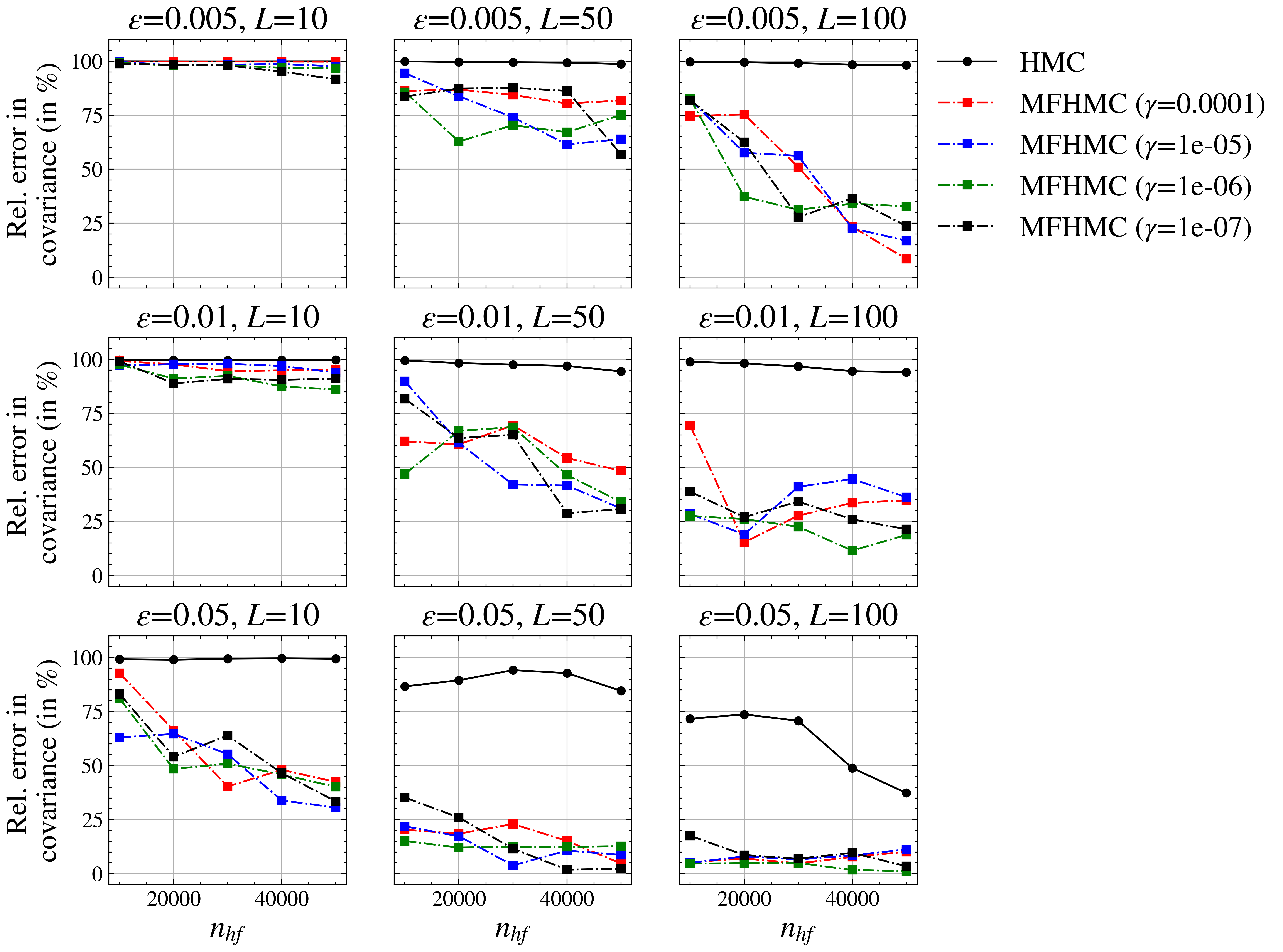}
    \caption{Comparison of relative error in covariance (in \%) for HMC and MFHMC at different computation budgets (as defined by the number of HF target distribution evaluations ($n_{hf}$)). To minimize the influence of randomness, each experiment was performed with five different random seeds. Each data point in the figure represents the average value of these five experiments.}
    \label{fig:cov_vs_hf_250mvn}
\end{figure}

Figures \ref{fig:ar_vs_hf_250mvn} and \ref{fig:ess_vs_hf_250mvn} present a relative comparison between the HMC and MFHMC algorithms (utilizing different LF models) concerning both computational and statistical efficiency, while \cref{fig:cov_vs_hf_250mvn} provides a comparison of their accuracy. These comparisons are conducted across various trajectory lengths ($\epsilon\times\ L$) and fixed computation budgets defined by the number of HF target density (\cref{eq:mult_gaussian_hf}) evaluations ($n_{hf}$), ranging from $n_{hf}=[1\times10^4, 2\times10^4, 3\times10^4, 4\times10^4, 5\times10^4]$.

From these figures, it can be observed that for very small trajectory lengths, both HMC and MFHMC make very small jumps. Consequently, they can only explore a small region of the parameter space within the given fixed maximum compute budget of $n_{hf}=5\times10^4$, leading to high errors in posterior covariance estimates. Furthermore, the samples exhibit high correlation due to these smaller jumps, resulting in lower values of ESS/$n_{hf}$ for both methods (with slightly better ESS/$n_{hf}$ for MFHMC compared to HMC).

However, the distinction between the two algorithms becomes evident when we increase the values of $\epsilon$ and $L$ to reasonable trajectory lengths. For these longer trajectories, both algorithms make substantial jumps in the parameter space. However, with HMC, most samples are rejected due to the sampler often venturing into low-probability regions of the parameter space with such large jumps. This results in a very small number of accepted moves/$n_{hf}$ value across all compute budgets. Posterior covariance statistics computed with such few samples naturally lead to very high errors, as depicted in \cref{fig:cov_vs_hf_250mvn}.

In contrast, MFHMC does not face these challenges with long trajectory lengths due to its two-stage structure. Consider a scenario where a jump leads into a low-probability region. The first (LF) stage of MFHMC will likely reject such samples, reducing the need for expensive HF evaluations for such ``bad'' samples. Consequently, most samples passed to the second stage for HF evaluation are likely in high-probability regions and are eventually accepted (provided the LF model is sufficiently close to the HF model). This results in a high number of accepted moves per $n_{hf}$ for MFHMC, as shown in \cref{fig:ar_vs_hf_250mvn}.

Furthermore, since MFHMC achieves a high acceptance rate for the second stage, it generates many more accepted samples for any given fixed compute budget. This improves the approximation of the target posterior significantly, leading to a substantial reduction in the error of posterior covariance estimates compared to HMC ( \cref{fig:cov_vs_hf_250mvn}). Additionally, as the samples accepted by MFHMC are less correlated, it yields a significantly higher ESS per $n_{hf}$ value across all compute budgets compared to HMC, almost an order of magnitude better.

This underscores an important feature of MFHMC: it achieves significantly better ESS and acceptance rates simultaneously while selecting ($\epsilon, L$) for longer jumps, eliminating the trade-off faced with traditional HMC between achieving a better acceptance rate (with shorter jumps) or a better effective sample size (with longer jumps)

Another intriguing observation from these figures is that in MFHMC, as the fidelity of the LF model increases (moving from $\gamma=1e-4$ to $1e-7$), the overall performance of MFHMC improves. This improvement is reflected in higher accepted moves per $n_{hf}$, higher ESS per $n_{hf}$, and reduced error in the posterior. This observation is intuitive since a more accurate LF model better mimics the behavior of the HF model. Consequently, it does a better job of filtering out bad samples in the first stage and passing good samples to the second stage, increasing the likelihood of acceptance by the HF model. This underscores the importance of using an accurate LF surrogate model in MFHMC.

\subsection{Initial condition inversion}\label{sec:init_cond_inv}
We next consider the problem of inferring the initial condition for the transient heat conduction problem given the noisy temperature measurement at some later time. 
\begin{align}\label{eq:transient_heat_cond}
    \nabla \cdot (\alpha \nabla u(\bm{s}, t)) &= \partial u(\bm{s}, t)/\partial t \qquad & \bm{s} \in \Omega, t \in (0, T] \\
    u(\bm{s}, 0) &= m(\bm{s}) \qquad & \bm{s} \in \Omega \\
    u(\bm{s}, t) &= g(\bm{s}, t) \qquad & \bm{s} \in \partial\Omega, t \in (0, T]
\end{align}
where $\Omega \subset \mathbb{R}^2$ is a square domain with length = $2\pi$ units and $\alpha=0.64$ unit is thermal diffusivity. $u(\bm{s}, t)$ is temperature at location $\bm{s}$ at time t, and $m(\bm{s})$ is the initial condition for temperature. Here, the parameter to infer $\infv$ is the nodal values of initial condition $m(\bm{s})$ and the measurement $\bm{y}$ is the nodal values of final temperature $u(\bm{s}, T)$.
The corresponding inverse problem is given noisy temperature-field measurements $\nmeas$ at later time $t=T=1$ unit infer the posterior distribution corresponding to the initial condition of temperature $\infv$. This is an ill-posed problem as significant information is lost via diffusion as we move forward in time.

We choose the second-order finite difference scheme in the spatial domain and backward Euler scheme in the temporal domain to solve the forward problem in \eqref{eq:transient_heat_cond} numerically. This leads to the following linear forward problem relating the inferred field (initial condition, $m(\bm{s})$) to the measured field (final temperature, $u(\bm{s}, T)$)
\begin{equation}\label{eq:forward_map}
    \bm{y} = \bm{F}\bm{x},
\end{equation}
where $\infv$ is a vector of nodal values of initial condition $m(\bm{s})$ and $\bm{y}$ is a vector of nodal values of final temperature field $u(\bm{s}, T)$. For our numerical experiment, we discretize the spatial domain in 32 nodal points in each direction and the temporal domain in 100 time steps. This will act as our forward model for HMC and the HF model for MFHMC.

In Bayesian inversion, we are interested in inferring the posterior density 
$$
\postinfv(\infv | \nmeas) \propto \plike(\nmeas|\infv)\priorinfv(\infv)
$$
To test the universality of the proposed algorithm, we consider two different prior distributions $\priorinfv(\infv)$: (i) Gaussian priors, and (ii) Generative Adversarial Network (GAN)-based priors\citep{patel2022solution}.

\subsubsection{Gaussian priors}\label{sec:init_cond_inv_gaussian_prior}\begin{wrapfigure}{r}{0.35\textwidth}
  \centering
  \includegraphics[width=0.32\textwidth]{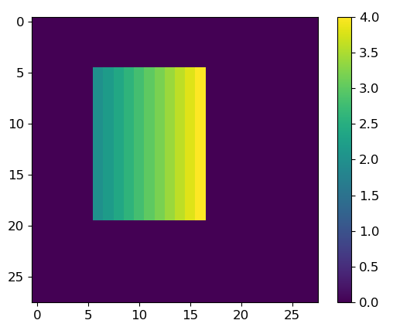}
  \caption{True initial condition ($\bm{x}$).}
  \label{fig:inf_field}
\end{wrapfigure}
\begin{figure*}[htbp]
    \centering \includegraphics[width=\linewidth]{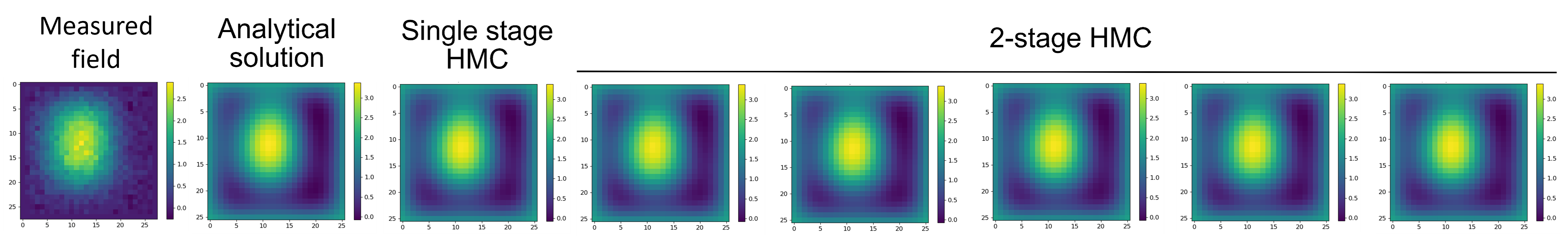}
    \caption{Qualitative comparison of the posterior mean for a Gaussian prior case. \textit{First column}: measured field. \textit{Second column} Analytical solution. \textit{Third column}: posterior mean computed using the single-stage HMC. \textit{4--8 column}: posterior mean computed using the proposed MFHMC algorithm with different surrogate models. Each of these models is based on truncated SVD with a different number of retained modes. Model 1: 25 modes, Model 2: 50 modes, Model 3: 75 modes, Model 4: 100 modes, Model 5: 200 modes.}
    \label{fig:stats_compare}
\end{figure*}

In this section, we select a Gaussian distribution for prior and likelihood distribution with zero mean and constant variance. Specifically, we select $\priorinfv(\infv) = \plike(\nmeas|\infv) = \mathcal{N}(\bm{0}, \sigma^2\bm{I_N})$ with $\sigma=0.1$. While selecting a Gaussian distribution for likelihood is relatively common, it is not common to select a Gaussian distribution for the prior for such problem. However, here, we select a Gaussian for a conjugate prior case allowing an analytical solution for the resulting posterior distribution. This, in turn, will allow us to compare the posterior QoIs obtained by our proposed two-stage HMC algorithm with the ``true'' QoIs. We again use the single-stage HMC algorithm as a baseline to compare our method's computational efficiency and accuracy. For this Gaussian prior case, we use a truncated singular value decomposition (TSVD) model as a surrogate ($\flf$) forward model. We consider five different surrogate forward models based on five different numbers of retained modes in TSVD: 25, 50, 75, 100, 200. 
\Cref{fig:inf_field} shows the ``true'' (discretized) initial condition field. Using this initial condition in Eq.\eqref{eq:forward_map}, the final temperature at time $t=T$ is obtained, and then uncorrelated Gaussian noise (with zero mean and 0.01 variance) is added to it (to avoid inverse crime) to obtain the synthetic measured temperature field (shown in \cref{fig:stats_compare} (leftmost panel)).


Next, we use this measured temperature field $\nmeas$  to obtain the posterior distribution $\postinfv(\infv|\nmeas)$ and compute its mean. We first compute the analytical solution for the mean for this conjugate prior case. This is shown in the second column of \cref{fig:stats_compare}. Next, we compute the posterior mean using the samples produced by the single-stage HMC algorithm (described in Section \ref{sec:background}). The results for this case are shown in the third column of \cref{fig:stats_compare}. Finally, we compute the posterior mean using the multi-fidelity HMC algorithm. We report results for all five surrogate models (model 1: 25 modes, model 2: 50 modes, model 3: 75 modes, model 4: 100 modes, model 5: 200 modes) in the last five columns of \cref{fig:stats_compare}. As can be observed from these subplots, the computed mean with the proposed method is in good qualitative agreement with the analytical solution as well as the solution obtained from the benchmark. Both single-stage and two-stage HMC algorithms were run for a fixed number of (20,000) MCMC steps, and the step size and the number of leapfrog steps were tuned to obtain the target acceptance ratio in the range of 0.6--0.7 following the recommendations of \citep{Beskos2013}.

\begin{table*}[bp]
\renewcommand{\arraystretch}{1}
\centering
\caption{Quantitative comparison of HMC and different MFHMC models (for Gaussian prior with TSVD-based surrogate model)}
\begin{adjustbox}{width=\textwidth}
\begin{tabular}{c c c c c c c}
\toprule[\heavyrulewidth]
\multirow{2}{*}{Model} & \multirow{2}{*}{Modes} & Acceptance rate & Acceptance rate & No. of HF & No. of rejected & Error in \\ & & (LF) & (HF) & evaluations & HF evaluations & mean (\%)  \\
\midrule[\heavyrulewidth]
Model 1 & 25 & 0.59 & 0.76 & 11845 & 2885 & 4.03  \\
Model 2 & 50 & 0.58 & 0.98 & 11664 & 213 & 3.47  \\
Model 3 & 75 & 0.59 & 0.99 & 11779 & 33 & 3.17  \\
Model 4 & 100 & 0.59 & 0.99 & 11720 & 5 & 3.13  \\
Model 5 & 200 & 0.59 & 1.0 & 11775 & 0 & 3.31  \\
\arrayrulecolor{black!30}\midrule
Model 2a & 50 & 0.02 & 0.98 & 445 & 5 & 3.65  \\
Model 5a & 200 & 0.02 & 1.0 & 447 & 0 & 3.33  \\
\arrayrulecolor{black!30}\midrule
1 stage HMC & 900 (all) & --- & 0.59 & 400000 & 163500 & 3.21  \\
\bottomrule
\end{tabular}\label{tab:quant_comparison}
\end{adjustbox}
\end{table*}

Next, we compare the proposed algorithm quantitatively with the single-stage HMC algorithm in \cref{tab:quant_comparison}. We can observe the following points from \cref{tab:quant_comparison}:
\begin{itemize}
    \item Significant improvement in the acceptance ratio can be achieved with the two-stage algorithm. This translates into a smaller number of \textit{wasted} HF simulations. This is possible because most of the ``bad'' samples are filtered out by the first stage (without needing expensive an HF simulation and thus achieving considerable computational saving). And most of the samples passed to the second stage by the surrogate model are already of high quality and so they are retained by the second stage with high probability. This is not the case with the single-stage HMC algorithm, where there is no such filtering mechanism, and hence, all samples are evaluated by the HF model.
    
    \item The number of HF evaluations required by the single-stage HMC algorithm is very high. This is due to the fact that each HMC step requires solving the forward and the adjoint problem, where each forward and adjoint problem requires $L$ (number of leapfrog steps) number of HF evaluations (i.e., $2L$  number of total HF evaluations for a single HMC step). Whereas, in our proposed MFHMC algorithm, only a single HF evaluation (in the second stage) is required for samples accepted in the first stage and no HF evaluation is required for samples rejected in the first stage.
    
    \item As the fidelity of the low fidelity (LF) model improves, the acceptance rate of the HF model improves dramatically. This is not surprising since as the fidelity of the LF model improves, it becomes more and more similar to the HF model and hence the majority of the samples accepted by the LF model are eventually accepted by the HF model boosting its acceptance rate and reducing \textit{wasted} HF evaluations.
    
    \item The error of our proposed method is similar (or better) to the single-stage HMC algorithm. Furthermore, the accuracy can further be increased by using a high-quality LF model.
    
    \item We did two additional experiments with Model 2 and Model 5 as LF model (denoted as Model 2a and Model 5a in \cref{tab:quant_comparison}), where we tuned the HMC hyper-parameters (step size and the number of leapfrog steps) to achieve a longer trajectory length for the Hamiltonian dynamics for the proposed algorithm. This translates to very low acceptance rate for the first stage and highly uncorrelated samples for this problem, resulting in orders of magnitude fewer HF evaluations required. Even with such fewer HF evaluations and not fully converged chains, MFHMC can achieve accuracy comparable to a single-stage HMC for this simple problem.
\end{itemize}

\subsubsection{GAN-based priors}\label{sec:init_cond_inv_gan_prior}
While the Gaussian prior is useful for verification purposes, it is not necessarily the best prior for the given inferred field (as can be observed by comparing the true inferred field in \cref{fig:inf_field} with the inferred means of \cref{fig:stats_compare}). Recently, various deep generative models have shown tremendous promise as prior for accurate inference in Bayesian inversion. A particularly promising model among them is the Generative Adversarial Network (GAN) \citep{Goodfellow2014}. The most appealing features of this model are (i) its ability to map high-dimensional data into a low-dimensional latent space, and (ii) the ability to learn and model complex probability distributions. By leveraging these two features, in recent years GANs have shown impressive results as priors in various domains such as geoscience, physics, and computer vision \citep{Mosser2019, Patel2021GAN}.

This section considers using GAN-based priors for the initial condition inversion problem. We consider two different surrogate forward models: (i) TSVD as before and (ii) a deep neural network that maps the initial condition $m(\bm{s})$ to the final temperature field $u(\bm{s}, T)$. Moreover, we consider the scenario where the compute budget (defined as the number of HF evaluations) is fixed and compare the relative performance. We use same step size ($\epsilon=0.01$) and number of leapfrog steps ($L=10$) for both HMC and MFHMC for all cases. To avoid the influence of randomness on final comparision, we ran all experiments with five different random seeds and report the average results in all cases. The detailed architectures of the generator and the discriminator used for GAN-prior are shown in~\cref{fig:arch_init_cond}.

\begin{figure}[htbp]
    \centering \includegraphics[width=0.8\linewidth]{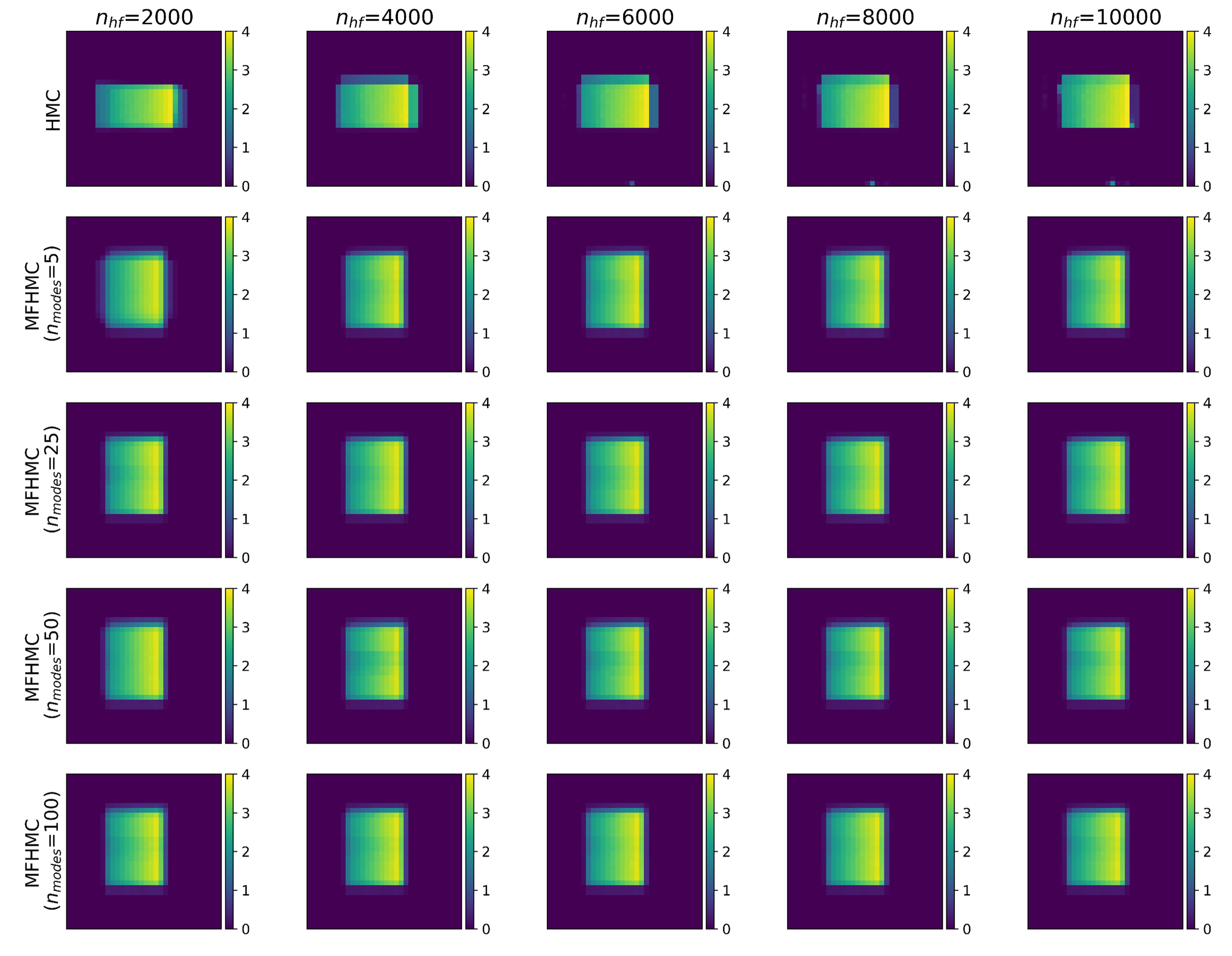}
    \caption{Qualitative comparison of posterior mean for initial condition inversion with GAN-based prior and TSVD-based LF surrogate models. \textit{Top row}: posterior mean obtained using HMC at different numbers of HF evaluations. \textit{Second to fifth row}: posterior mean obtained using different LF surrogate models ($n_{modes}=5, 25, 50, 100$ respectively) at different compute budget.}
    \label{fig:init_cond_gan_tsvd_qual}
\end{figure}

\paragraph{TSVD-based surrogate}
 \Cref{fig:init_cond_gan_tsvd_qual} shows results for Bayesian inversion with GAN-based prior and TSVD-based surrogate model for different compute budgets (as defined by the number of high fidelity evaluations $n_{hf}$) and different fidelities of the surrogate model (as defined by the number of retained modes $n_{modes}$). As a benchmark comparison, we again consider a single-stage HMC algorithm with HF forward model as described in Eq.\eqref{eq:forward_map}. As evidenced by the qualitative comparison in \cref{fig:init_cond_gan_tsvd_qual}, it is discernible that the posterior mean obtained through the employment of MFHMC with all surrogate models significantly better approximates the true inferred field ( \cref{fig:inf_field}) compared to HMC within an equivalent computational budget. The same trend continues at different numbers of high-fidelity evaluations. This observation underscores the efficacy of a two-stage algorithm in consistently and reliably estimating posterior statistics. 

\begin{figure}[htbp]
    \centering
    \includegraphics[width=0.8\linewidth]{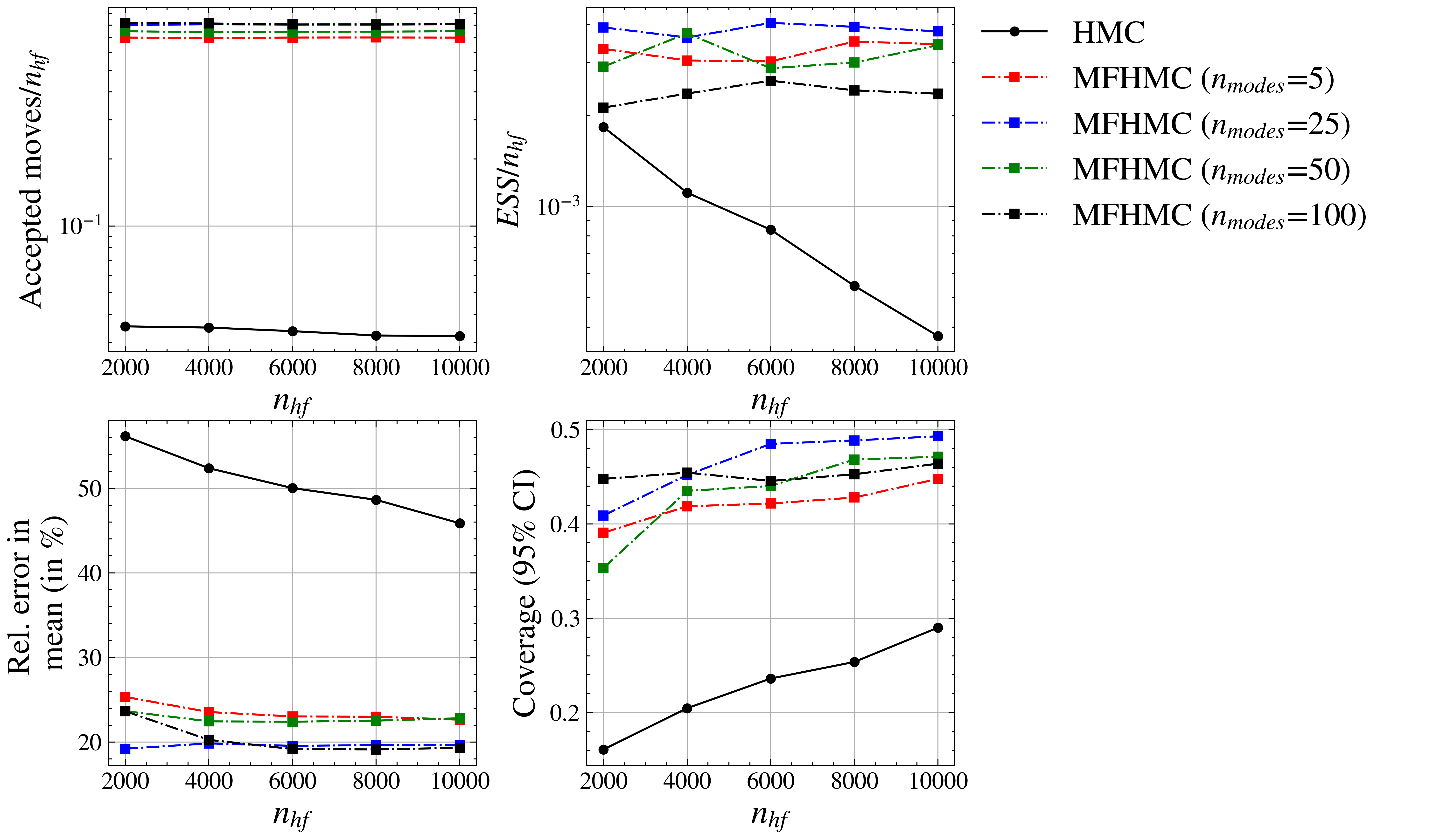}
    \caption{Quantitative comparison of the initial condition inversion problem using GAN-based prior and TSVD-based LF surrogate models. To minimize the influence of randomness, each experiment was performed with five different random seeds. Each data point in the figure represents the average value of these five experiments.}
    \label{fig:init_co}
\end{figure}

Detailed quantitative comparison of the same case is depicted in \cref{fig:init_co}. As can be observed from \cref{fig:init_co}, MFHMC (with all different surrogate models) has almost an order of magnitude higher $\text{ESS}/n_{hf}$ and Accepted moves/$n_{hf}$ compared to HMC at all computation budgets. \Cref{fig:init_co} also highlights that this better computational and statistical efficiency of MFHMC is not coming at the cost of accuracy, as it consistently shows lower relative error in posterior mean compared to HMC at all computation budgets. It is remarkable to note that even with 10,000 forward solves HMC has a higher error compared to any MFHMC model at 2,000 forward solves (as can also be observed from \cref{fig:init_cond_gan_tsvd_qual}). Such a virtuous combination of better efficiency and accuracy of MFHMC is due to its two-stage nature, which allows taking bigger jumps (resulting in higher $\text{ESS}/n_{hf}$), while only accepting better quality samples with its first stage filtering mechanism (resulting in lower relative error in posterior statistics). This also entails that most of the samples passed to the second stage (where only $\fhf$ comes into picture for MFHMC) are accepted (resulting in higher Accepted moves/$n_{hf}$). \Cref{fig:init_co} also shows that MFHMC (with all surrogate models) has significantly better uncertainty coverage than HMC at all computation budgets demostrating its superiority for uncertainty quantification. (Note: Coverage  is defined as the ratio of the number of nodes (pixels) where the true value lies within the 95\% confidence interval (CI) determined by the posterior mean and standard deviation to the total number of nodes). It can be observed from \cref{fig:init_co} that as the accuracy of the LF model increases (i.e., $n_{\text{modes}}$ increases), the relative performance of MFHMC also improves, which again is not surprising. 


\paragraph{DNN-based surrogate}
Up to now, we have considered TSVD-based surrogate forward models for which the gradient (in the first stage of the MFHMC algorithm) can be computed analytically. While this is an ideal surrogate model that shows excellent results for the linear initial condition inversion problem considered here, a TSVD-based surrogate model might not be sufficient for more complex and nonlinear forward problems due to its linear nature. In recent years, DNNs have emerged as powerful tools for building surrogates for highly nonlinear and complex forward models \citep{Zhu2019}. Moreover, the automatic differentiation capabilities of modern machine learning libraries enable gradient calculations with a high degree of accuracy at almost no additional cost. This feature is particularly useful in the first stage of the MFHMC algorithm. We test the effectiveness of such a DNN-based surrogate forward model in our proposed algorithm for the initial condition inversion problem described earlier in a controlled experiment.

\begin{figure}[htbp]
    \centering \includegraphics[width=0.6\linewidth]{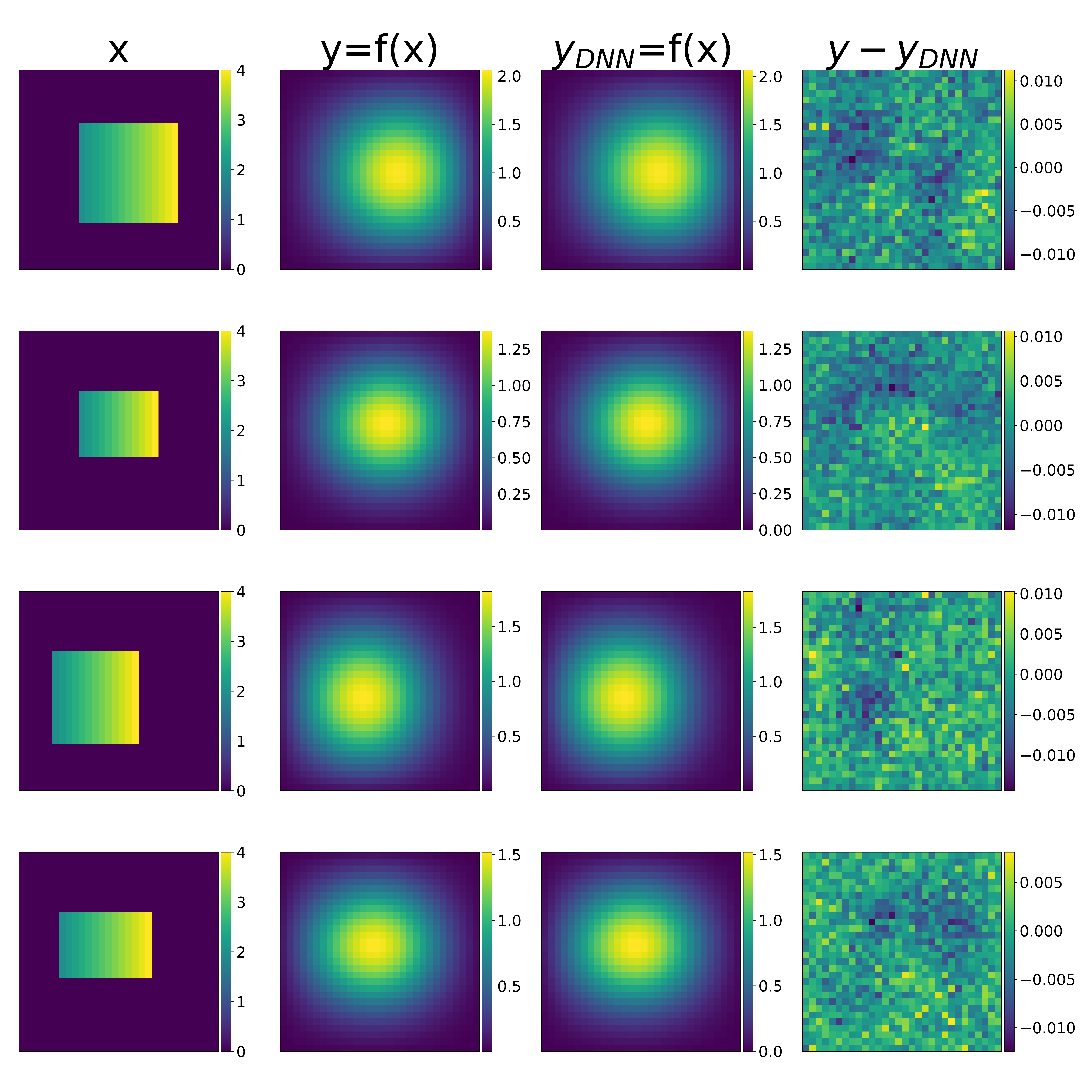}
    \caption{DNN-based surrogate model predictions for four different samples from the validation set. \textit{First column}: initial condition samples from the validation set. \textit{Second column}: true final temperature field obtained using the HF numerical solver. \textit{Third column}: final temperature field prediction by the trained DNN-based surrogate. \textit{Fourth column}: difference between the true and predicted temperature fields.}
    \label{fig:dnn_surrogate}
\end{figure}

For this, we first train a 4-layer deep fully connected neural network (with ReLU activation function in all layers except the last layer) as a surrogate ($\bm{f_{DNN}}$) mapping the vector of the initial condition field ($\infv$) to the vector of the final temperature field ($\bm{y}$) in \textit{an offline stage}. This DNN was trained by minimizing the mean square error loss using the Adam optimizer \citep{Kingma2014} with a batch size of 1024 and learning rate of 0.001 over $2\times10^3$ pair-wise training data (of the initial condition and final temperature generated using the HF numerical solver) for 1000 epochs. \Cref{fig:dnn_surrogate} shows the final temperature prediction of the trained DNN for four initial conditions sampled from the validation set. The relative error (measured in the $L_2$ sense) of the trained DNN on the validation set is $0.55\%$.

\begin{figure}[htbp]
    \centering \includegraphics[width=0.85\linewidth]{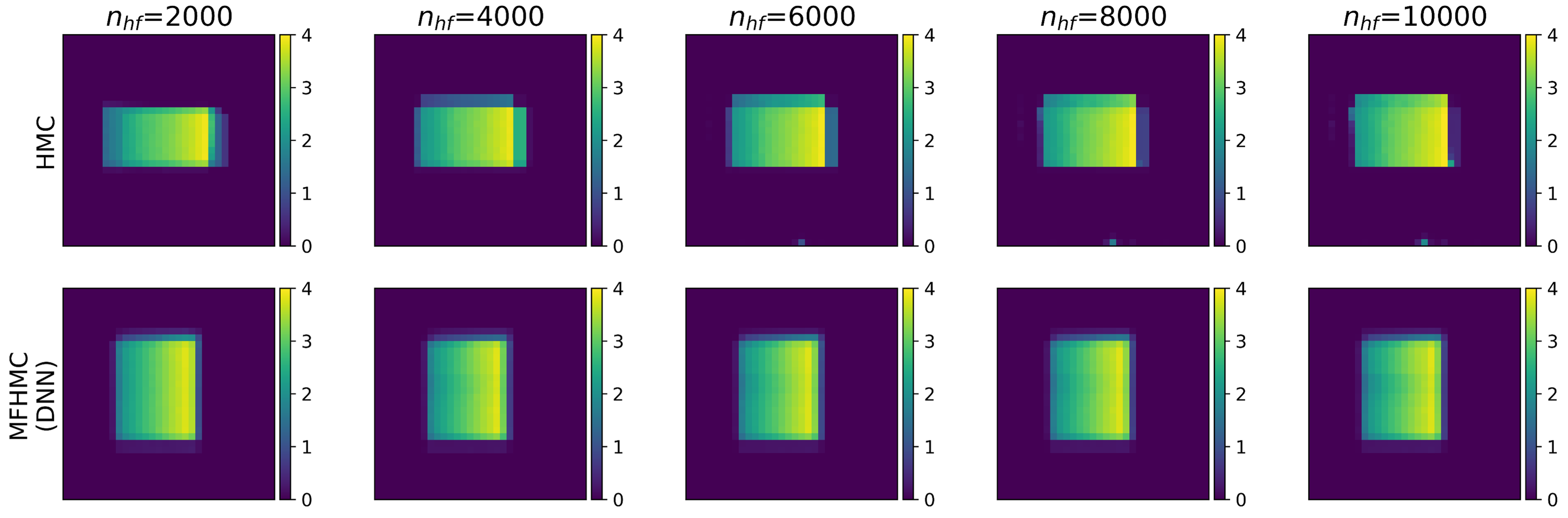}
    \caption{Qualitative comparison of posterior mean for initial condition inversion with GAN-based prior and DNN-based LF surrogate models. \textit{Top row}: posterior mean obtained using HMC at different numbers of HF evaluations. \textit{Second to fifth row}: posterior mean obtained using different LF surrogate models ($n_{modes}=5, 25, 50, 100$ respectively) at different compute budget.}
    \label{fig:init_cond_gan_dnn8k_qual}
\end{figure}

\begin{figure}[htbp]
    \centering
    \includegraphics[width=0.8\linewidth]{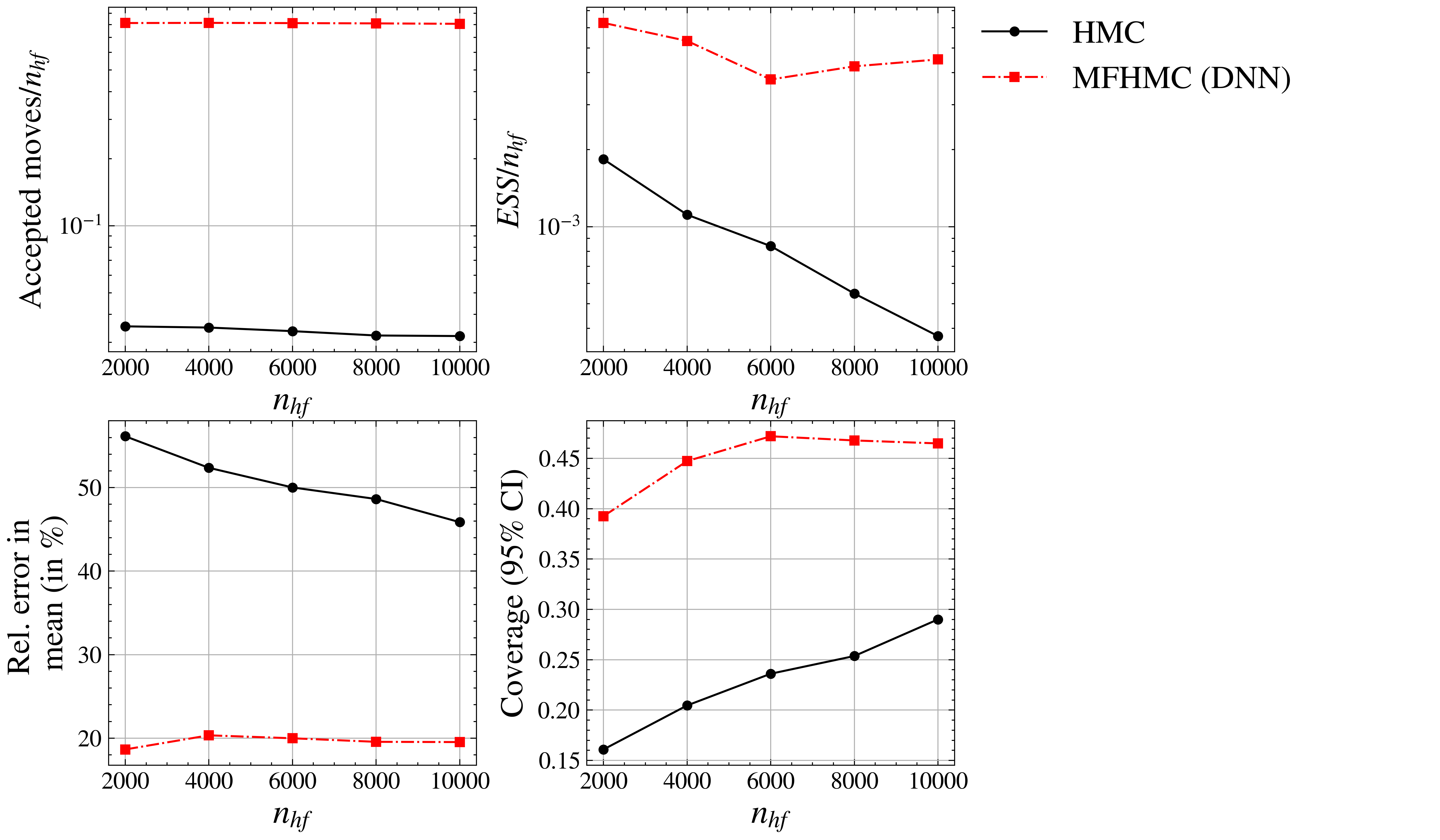}
    \caption{Quantitative comparison of the initial condition inversion problem using GAN-based prior and DNN-based LF surrogate models. To minimize the influence of randomness, each experiment was performed with five different random seeds. Each data point in the figure represents the average value of these five experiments.}
    \label{fig:init_cond_gan_dnn8k_qoi}
\end{figure}

Once the DNN is trained in an offline stage, it is then used as a surrogate forward model in the likelihood term of the first stage of the MFHMC method, and the gradient of the resulting posterior is computed using automatic differentiation. In the second stage of the MFHMC algorithm, we use the HF numerical solver as before.  \Cref{fig:init_cond_gan_dnn8k_qual} shows the qualitative comparison of the posterior mean for HMC and MFHMC with this DNN-based surrogate for different compute budgets. As can be observed from this figure, the posterior mean obtained by MFHMC significantly better approximates the true inferred field (\cref{fig:inf_field}) compared to HMC within an equivalent computational budget.  
The same observations can be made from \cref{fig:init_cond_gan_dnn8k_qoi}, which shows a quantitative comparison of HMC and MFHMC for the same case. In the bottom-left subplot, it is evident that the relative error in the posterior mean of HMC is significantly higher than that of MFHMC at all computational budgets. Moreover, the number of Accepted moves/$n_{hf}$ and $\text{ESS}/n_{hf}$ (shown in the top subplots) for MFHMC is again almost an order of magnitude higher than for HMC at all computational budgets, indicating its superior computational and statistical efficiency. Similarly, the coverage of MFHMC (bottom-right subplot) is significantly better than that of HMC at all computational budgets. Note that the number of HF solutions used for training the DNN-based surrogate is not included in \cref{fig:init_cond_gan_dnn8k_qoi}, as training is a one-time offline cost that can be amortized over multiple inferences. However, even if we include that training cost ($n_{hf}=2000$), MFHMC still significantly outperforms HMC at all compute budget in all metrics. This emphasizes the highly promising potential of MFHMC, showcasing its ability to potentially replace HMC even in a white-box setting due to its superior computational and statistical efficiency and accuracy, even when factoring in the training cost. We anticipate that with the ongoing advancements in accurate deep learning-based surrogates \citep{lu2021learning, li2020fourier, patel2024variationally}, MFHMC could become a preferred choice over HMC across various applications in the future.

\subsection{Coefficient inversion}\label{sec:coefficient_inv}
This is a non-linear coefficient inversion problem for an elliptic PDE which arises in many fields such as subsurface flow modeling \citep{Iglesias2013EvaluationOG}, electrical impedance tomography \citep{kaipio2006statistical}, and inverse heat conduction \citep{Kaipio2011}. Here, the goal is to infer the coefficient of the PDE (permeability/thermal conductivity field) given noisy measurements of the pressure/temperature field. For this problem, the forward model is described as:
\begin{equation}\label{eq:conductivity}
\begin{aligned}
-\nabla \cdot (\kappa (\bm{s}) \nabla u (\bm{s})) &= b(\bm{s}), & \qquad & \bm{s} = (s_1, s_2) \in \Omega & \\
u (\bm{s}) &= 0, & \qquad &\bm{s} = (s_1, s_2) \in \partial \Omega &
\end{aligned}
\end{equation}
where $\Omega \subset \mathbb{R}^2$ is a square domain with length = 1 unit, and $b(\bm{s}) = 10^3$ units denotes the heat source. The goal is to infer the posterior QoIs of permeability/conductivity field $\kappa$, given a noisy, and potentially partial, measurement of pressure/temperature field $u$. The nodal values of the pressure/temperature field are stored in the vector $\bm{y}$, and those of the permeability/conductivity field are stored in the vector $\bm{x}$. 

In this experiment, we consider two different datasets: a parametric (rectangular) dataset and a channelized flow dataset. The first dataset corresponds to the inverse heat conduction problem, while the second corresponds to the permeability inversion problem commonly encountered in geophysics. We further evaluate two separate scenarios for each dataset to assess their performance. For the rectangular dataset (inverse heat conduction problem), we explore the scenario with a fixed number of MCMC steps and for the channelized flow dataset (permeability inversion problem), we examine a scenario with a fixed computational budget for HF evaluations. GAN-based priors are considered for both datasets. The detailed architectures of the generator and the discriminator used for GAN-prior are shown in~\cref{fig:arch_channelized_flow}.


\subsubsection{Inverse heat conduction}\label{sec:darcy_rectangle}

\begin{figure}[htbp]
    \centering \includegraphics[width=0.85\linewidth]{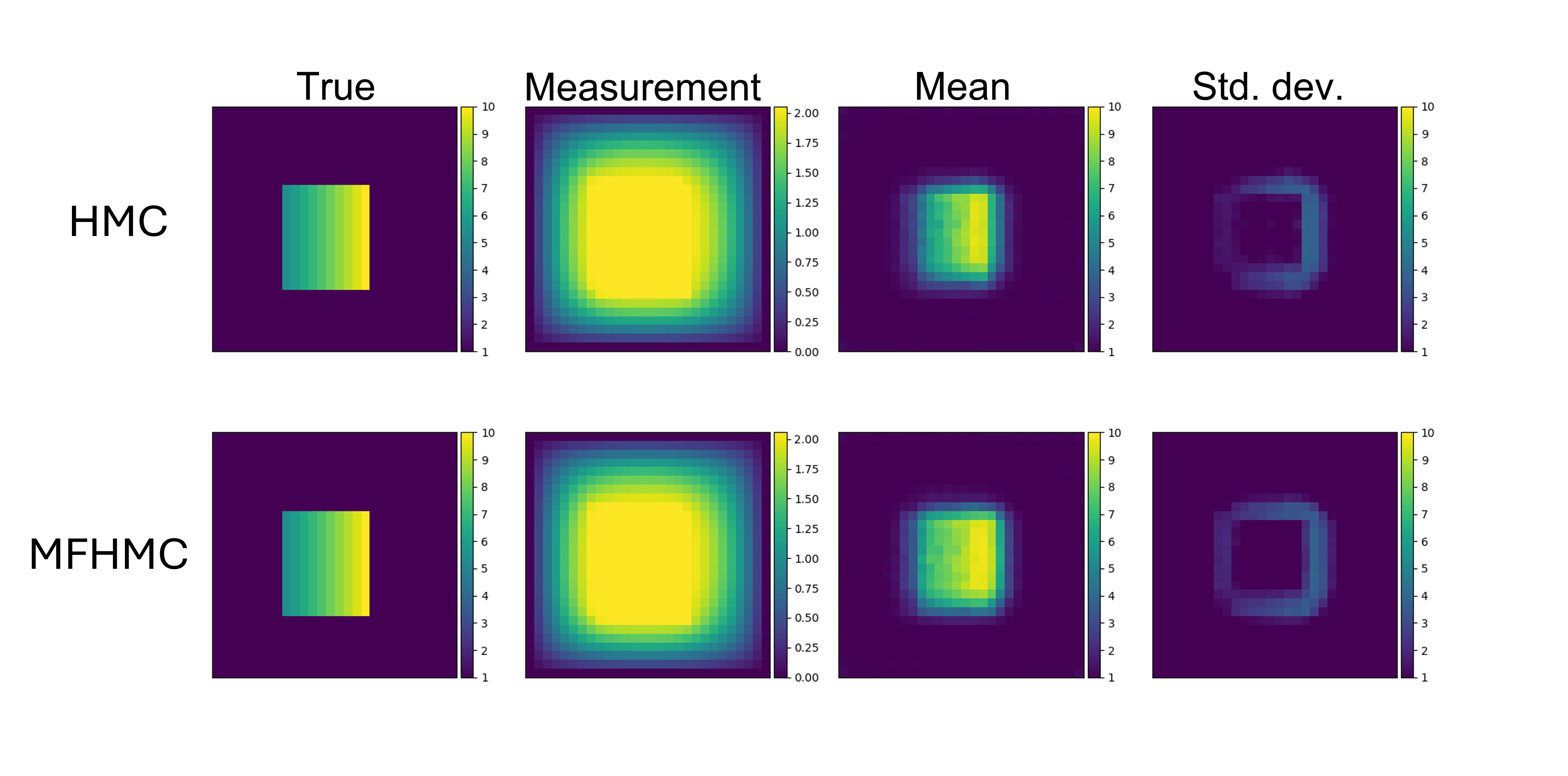}
    \caption{Comparison of the posterior QoIs for the proposed MFHMC algorithm (\textit{second row}) with the single stage HMC algorithm (\textit{first row}) for the fixed statistical error scenario.}
    \label{fig:darcy_rectangle}
\end{figure}

In this experiment, we utilize the parametric rectangular dataset discussed in the previous section and used in earlier studies for similar inverse problems~\citep{dasgupta2024dimension}. We use a convolutional neural network (CNN)-based surrogate as LF and use finite element solution of~\cref{eq:conductivity} with linear finite elements as HF model. The detailed architecture of the LF model is provided in \cref{fig:arch_channelized_flow}. We examine the scenario of a fixed number of MCMC steps  (i.e., approximately same statistical error in the posterior statistics for both algorithms) to compare the relative performance of both. \Cref{fig:darcy_rectangle} shows the posterior statistics for both algorithms. As observed in this figure, the posterior statistics are more or less the same for both algorithms, but MFHMC needs significantly fewer HF evaluations (as indicated in \cref{tab:darcy_rectangle}).


\begin{table}[htbp]
\renewcommand{\arraystretch}{0.9}
\centering
\caption{Comparison of HMC and MFHMC algorithm for the rectangular dataset}
\begin{adjustbox}{width=0.6\linewidth}
\begin{tabular}{c c c}
\toprule
Quantity of Interest & HMC & MFHMC  \\
\midrule
No.\ of HF evaluations ($n_{hf}$) $\downarrow$ & 13,450 & \textbf{4,797}\\

ESS/$n_{hf}$ $\uparrow$ & $\mathcal{O}(10^{-5})$  & $\bm{\mathcal{O}(10^{-4})}$\\

ESJD/$n_{hf}$ $\uparrow$ & $\mathcal{O}(10)$  & $\bm{\mathcal{O}(100)}$\\
\bottomrule
\end{tabular}\label{tab:darcy_rectangle}
\end{adjustbox}
\end{table}

Quantitative comparison of HMC and MFHMC is provided in \cref{tab:darcy_rectangle}. As can be observed from this table, MFHMC offers an order of magnitude improvement over HMC in normalized ESS and ESJD value while requiring significantly less HF evaluations.

\subsubsection{Permeability inversion}\label{sec:permeability_inversion}
\begin{figure}[htbp]
    \centering \includegraphics[width=0.85\linewidth]{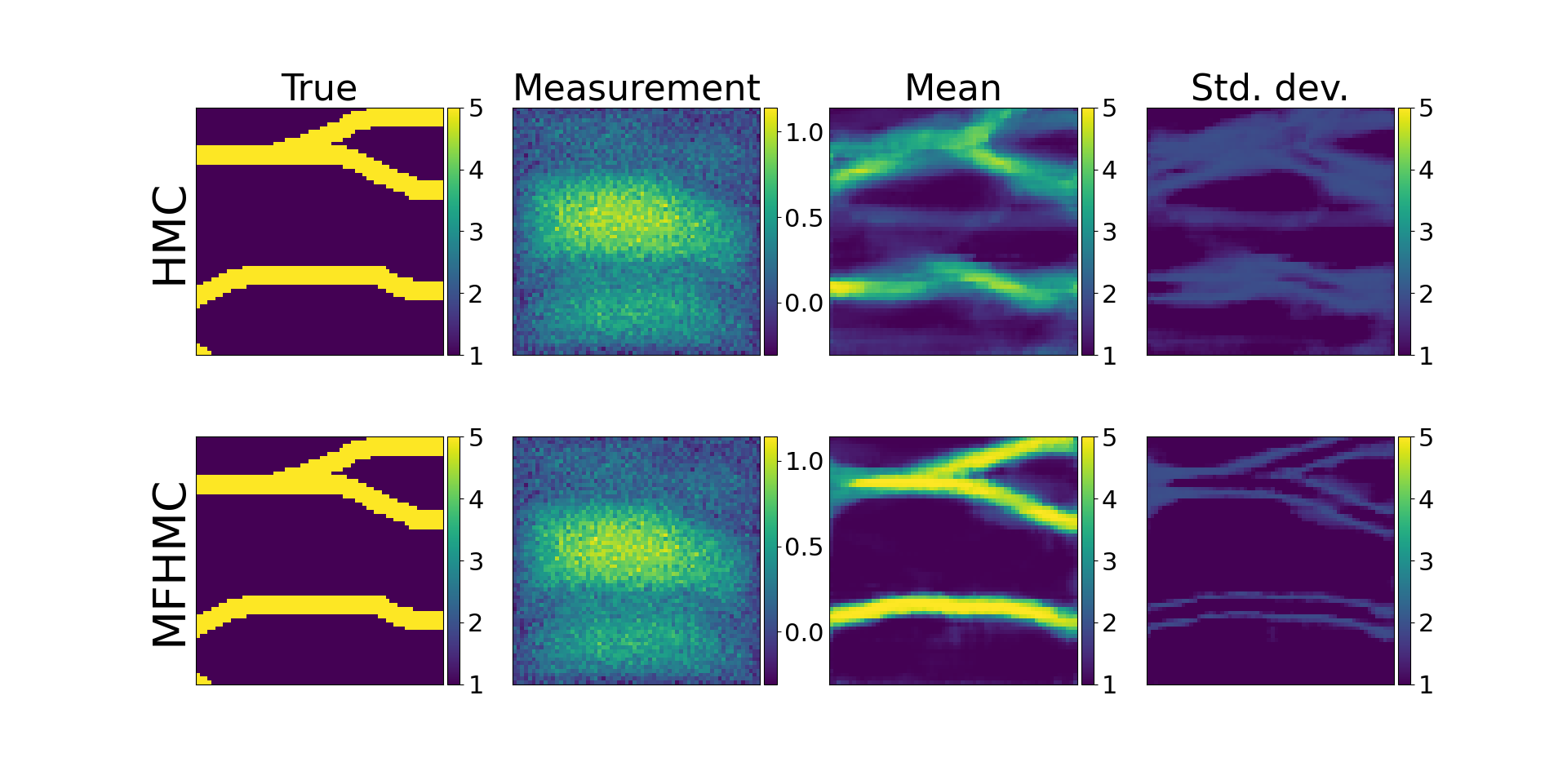}
    \caption{Comparison of the posterior QoIs for the proposed MFHMC algorithm (\textit{second row}) with the single stage HMC algorithm (\textit{first row}) for the fixed number of HF simulations.}
    \label{fig:channelized_flow_compare}
\end{figure}
For this experiment, we use a channelized flow dataset which is popular in geophysics. Again we use a CNN-based surrogate as LF model and use finite element solution of~\cref{eq:conductivity} with linear finite elements as HF model. The LF model was trained using 4,000 training samples (of pairwise data of permeability and pressure fields) using Adam optimizer and MSE loss function. The detailed architecture of the LF model is provided in \cref{fig:arch_channelized_flow}.
We further consider the scenario where we have a fixed computational budget of 10,000 HF simulations and ran single-stage HMC and MFHMC algorithms and computed posterior statistics such as mean and standard deviation.
\Cref{fig:channelized_flow_compare} shows these posterior statistics. As can be observed from these figures, compared to HMC, the mean of the proposed algorithm is much closer to the ground truth and the standard deviation plot captures the regions of uncertainty. In \cref{tab:channelized_flow}, we compare the two algorithms quantitatively. We consider the coverage and the error in the posterior mean as our evaluation metric. For error computation, we use the true value of the permeability field as our reference value.
\begin{table}[ht]
\renewcommand{\arraystretch}{0.9}
\centering
\caption{Comparison of HMC and MFHMC algorithm for the channelized flow dataset}
\begin{tabular}{c c c}
\toprule
Quantity of Interest & HMC & MFHMC  \\
\midrule
Error in mean (in \%) $\downarrow$ & 51.4 & \textbf{31.4}\\
Coverage (95\% CI) $\uparrow$ & 0.62  & \textbf{0.89}\\
\bottomrule
\end{tabular}\label{tab:channelized_flow}
\end{table}

\subsection{Laboratory-scale hydraulic tomography}\label{sec:real_tomography}
In this section, we demonstrate the effectiveness of our proposed algorithm with a real-world experimental dataset. Specifically, we consider a laboratory-scale hydraulic tomography experiment where the pressure/hydraulic head changes are recorded from a series of pumping tests in order to reconstruct the spatially distributed hydraulic conductivity field of a lab-scale sandbox. The experiments were conducted at the University of Iowa by Walter Illman and colleagues, and the same set of data have been used previously in various studies \citep{illman2007steady,Liu2011}.



\begin{figure}
    \centering
    \includegraphics[width=\linewidth]{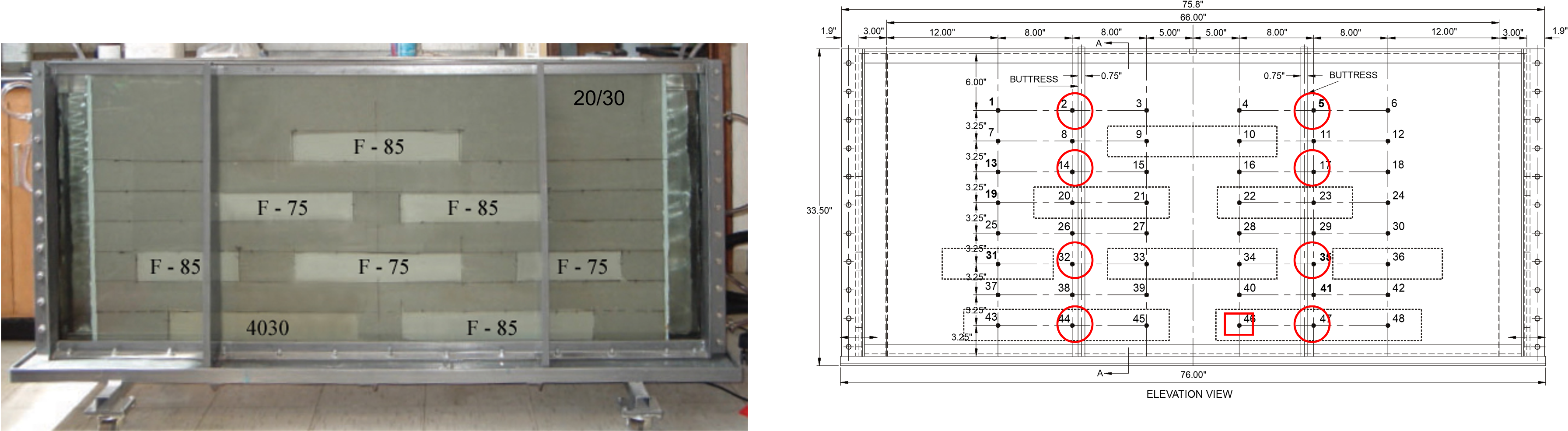}
    \caption{\textit{Left panel}: Front view of the Sandbox used in the hydraulic tomography experiment. \textit{Right panel}: Schematic of the sandbox with the dimensions of various sand blocks and the sensor locations. The numbered nodes (1--48) indicates the location of pressure sensor. The red circle indicates the location of source sensor. Figure reused from \citep{Liu2011}. The details of the sandbox flow-through tests can be found in \citet{illman2007steady}}
    \label{fig:tomo_exp_schematic}
\end{figure}

\begin{figure}[htbp]
    \centering \includegraphics[width=\linewidth]{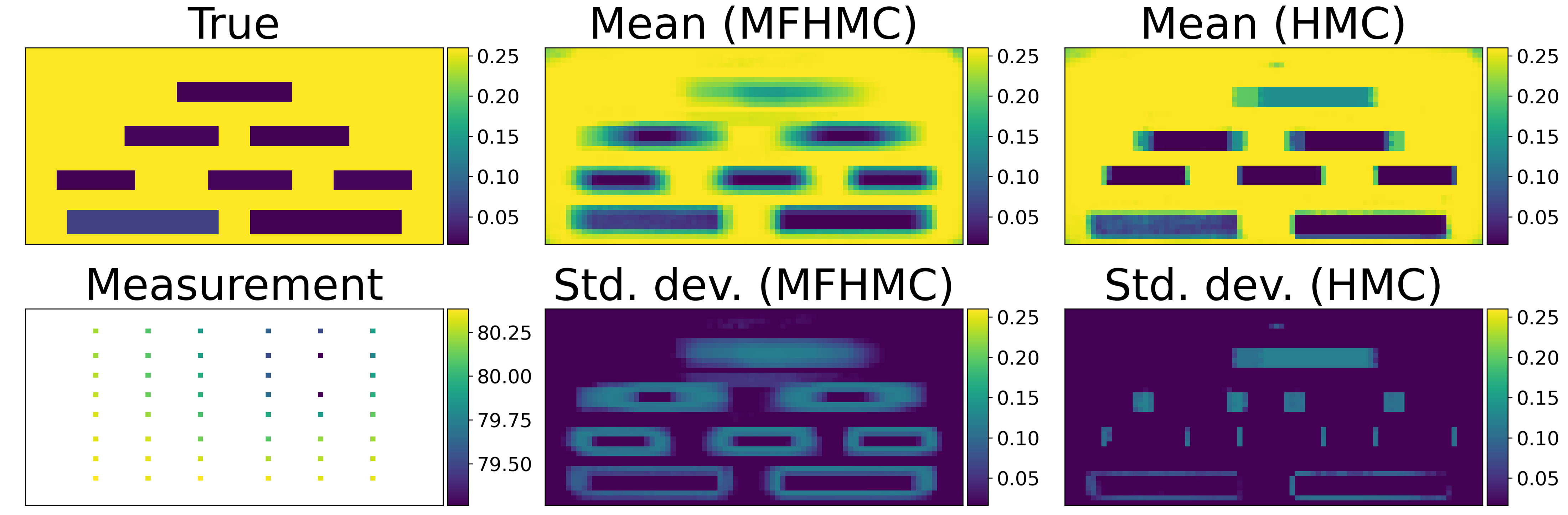}
    \caption{Inverting permeability for the hydraulic tomography problem. True hydraulic conductivity field from the experiment (top left panel). The measured pressure field values at the sensor nodes from one of the nine experiments (bottom left panel). The mean of the posterior distribution inferred using the MFHMC  (top middle panel) and HMC (bottom middle) algorithm respectively. The standard deviation of the posterior distribution inferred using the MFHMC (top right panel) and HMC (bottom right panel) respectively.}
    \label{fig:tomography_compare}
\end{figure}

\Cref{fig:tomo_exp_schematic} shows the front view (left panel) and schematic (right panel) of the sandbox used in this experiment, respectively. The dimensions of the sandbox are 161 cm long and 81 cm high. Four different kinds of commercially available sands (denoted by F-85, F-75, 4030, and 20/30 in \cref{fig:tomo_exp_schematic} (left panel)) were used to construct the sandbox. The box is composed of finer sand, and the background is made of coarser sand. In particular, eight rectangular slots (indicated by F-85, F-75, 4030) were packed with the sand with low permeability, whereas the background (20/30) was packed with sand with higher permeability. The true value of permeability at different locations is shown in \cref{fig:tomography_compare}. On the back of the sandbox 48 pressure sensors were installed at locations indicated by solid dots with numbers in \cref{fig:tomo_exp_schematic} (right panel) to measure the hydraulic head. Nine different experiments were conducted for hydraulic survey (at the locations indicated by red circled and squared ports in \cref{fig:tomo_exp_schematic} (right panel)) resulting in 9 x 47 = 423 measurements in total. Data from all nine experiments were used for inference.

The governing problem for this hydraulic survey is given by
\begin{equation}\label{eq:tomography}
\begin{aligned}
-\nabla \cdot (\kappa (\bm{s}) \nabla u (\bm{s})) &= f_i\delta(\bm{s}-\bm{a}), & \qquad & \bm{s} = (s_1, s_2) \in \Omega & \\
u (\bm{s}) &= g_i, & \qquad &\bm{s} = (s_1, s_2) \in \partial \Omega_g & \\
\kappa(\bm{s}) \nabla u (\bm{s}) &= h_i, & \qquad &\bm{s} = (s_1, s_2) \in \partial \Omega_h & \\
\end{aligned}
\end{equation}
where $\kappa$, $u$, and $\bm{a}$ indicate hydraulic conductivity, hydraulic head, and circled ports' locations, respectively. $f_i$, $g_i$, and $h_i$ indicate source, Dirichlet boundary condition, and the flux boundary condition for $i^{th}$ experiment with $i=\{1, 2, \dots, 9\}$. 
\begin{figure}[htbp]
    \centering \includegraphics[width=0.85\linewidth]{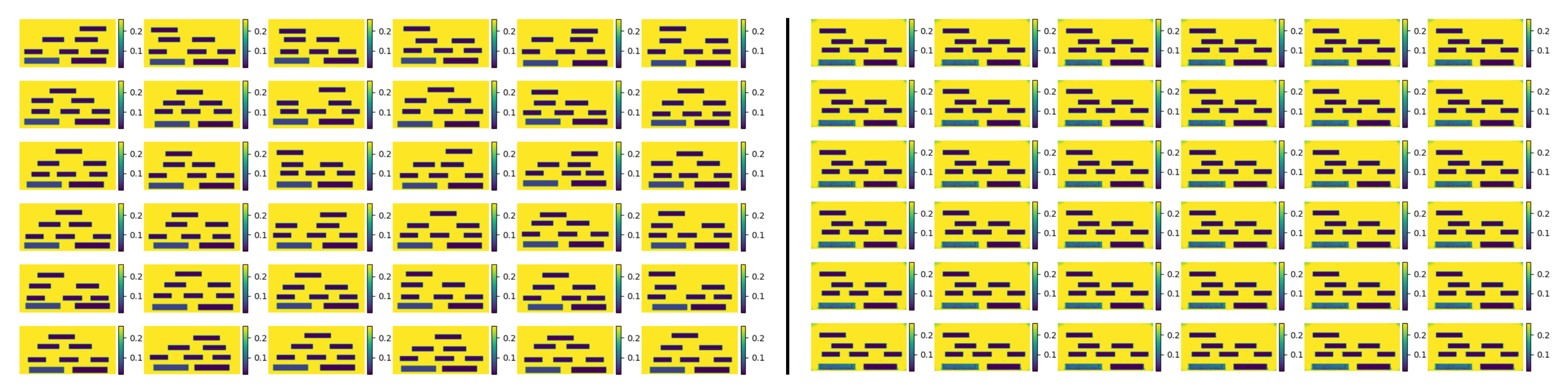}
    \caption{Thirty six realizations of the hydraulic conductivity from the training set used to train the GAN-prior (left panel) and corresponding realizations generated by the trained GAN-prior (right panel).}
    \label{fig:gan_tomography}
\end{figure}

In order to perform the inference, we use the GAN-based priors. The detailed architectures of the generator and the discriminator are provided in \cref{fig:arch_real_tomo}. This prior was trained by varying the horizontal and vertical location of the eight sand blocks. \Cref{fig:gan_tomography} shows the realizations of hydraulic tomography from the training set used to train the GAN-prior and corresponding realizations from the trained GAN. For the LF model, a CNN-based surrogate model was used to map the hydraulic conductivity image to 432 hydraulic head measurements corresponding the 47 head measurements for all nine experiments. The architecure of this surrogate is provided in ~\cref{fig:arch_real_tomo}. For the HF model, a finite element solution of \cref{eq:tomography} was used through FEniCS \citep{AlnaesEtal2015}. Again, we use HMC as a baseline method for comparison. We run both HMC and MFHMC for the fixed compute budget of $n_{hf}=1000$ with step size and number of leapfrog steps set to 0.01 and 10 respectively.
\Cref{fig:tomography_compare} shows the posterior distribution's mean and standard deviation inferred using the MFHMC and HMC algorithm. As can be observed, the mean estimate of MFHMC is able to capture the location of all the boxes quite well except for the top sandbox. It can also accurately infer the value of hydraulic conductivity for all the sandboxes except the top one. We hypothesize the difficulty in capturing the exact value of hydraulic conductivity for this box is probably due to the noisy measurement from the top sensors. In comparison HMC seems to have sharper and slightly erroneous mean estimate. This can be observed quantitatively from \cref{tab:tomo} with HMC's higher error in mean estimate. The bottom middle panel in \cref{fig:tomography_compare} shows the standard deviation of the posterior distribution obtained using MFHMC algorithm. This standard deviation field, which is our measure of uncertainty, is elevated along the edges of the sandboxes. This is expected as perturbing the location of the sandbox slightly will not have any significant effect on measurements and hence leads to high uncertainty \citep[e.g.,][]{lee2013bayesian}. Moreover, uncertainty is highest for the top sandbox where mean estimate fails to accurately capture the true field, which aligns well with the intuition. In comparison HMC's uncertainty estimates are not quite accurate as also depicted by its lower coverage in \cref{tab:tomo}. Finally, MFHMC has two orders of magnitude higher number of accepted moves/$n_{hf}$ compared to HMC for this specific problem. Better reconstruction results along with its superior efficiency highlight MFHMC's relative effectiveness in dealing with such noisy experimental data. 
\begin{table}[ht]
\renewcommand{\arraystretch}{0.9}
\centering
\caption{Comparison of HMC and MFHMC algorithm for the hydraulic tomography problem}
\begin{tabular}{c c c}
\toprule
Quantity of Interest & HMC & MFHMC  \\
\midrule
Error in mean (in \%) $\downarrow$ & 34.8 & \textbf{26.9} \\
Coverage (95\% CI) $\uparrow$ & 0.30  & \textbf{0.77} \\
Accepted moves/$n_{hf}$ $\uparrow$ & $\mathcal{O}(10^{-3})$ & $\bm{\mathcal{O}(10^{-1})}$\\ 
\bottomrule
\end{tabular}\label{tab:tomo}
\end{table}


\section{Conclusion}\label{sec:conclusion}
HMC is a powerful method for generating samples from an unnormalized probability distribution. By exploiting the geometry of the target density, it can achieve a faster convergence rate than traditional MCMC methods and is well-suited for scaling to high-dimensional parameter spaces. However, using an HMC algorithm for many PDE-driven Bayesian inverse problems is infeasible or computationally impractical due to the black-box nature of many forward model simulators and/or expensive gradient computations.

In this manuscript, we have proposed a novel “gradient-free” HMC algorithm, which does not require gradient of the posterior and as a result is compatible with black-box simulators and non-differentiable priors while enjoying the favorable properties of HMC (faster convergence, scalability to high dimensions). The effectiveness and versatility of the proposed algorithm is demonstrated through a series of numerical experiments involving both simulated and experimental data as well as variety of forward model surrogates. All numerical experiments indicate the method performs superior to the traditional HMC algorithm in terms of both computational and statistical efficiency (sometimes by  orders of magnitude)  while maintaining high accuracy. Furthermore, the proposed algorithm appears more robust to the choice of hyperparameters (number of leapfrog steps and step size) compared to HMC. It also enables a longer trajectory length (larger jumps) and a higher acceptance rate simultaneously.

There are several interesting avenues for further study in this area. On the theoretical front, investigating the optimal acceptance rate for MFHMC (similar to that for HMC~\citep{Beskos2013}) and its dependence on the accuracy of the surrogate model used in the first stage would be valuable. On the methodological side, extending the proposed multi-fidelity strategy to advanced versions of HMC algorithms such as Riemannian Manifold Hamiltonian Monte Carlo (RMHMC)\citep{girolami2011riemann} (a variant of HMC that simulates Hamiltonian dynamics in Riemannian rather than Euclidean spaces) or No-U Turn Sampler (NUTS)\citep{hoffman2014no} would be of particular interest. Another intriguing research direction of potentially considerable practical interest could involve designing algorithms tailored to the availability of surrogate models. For instance, while this manuscript assumes access to one high-fidelity and one low-fidelity forward model, many computational science and engineering applications involve hierarchies of low-fidelity models~\citep{geraci2017multifidelity}. Extending the current MFHMC algorithm to such multi-level multi-fidelity setting could prove useful. Conversely, in scenarios lacking obvious computationally cheap surrogate models~\citep{sherlock2017adaptive}, developing an adaptive MFHMC algorithm that can build an approximation of the posterior in the first stage adaptively based on high-fidelity posterior evaluations would be interesting.

\appendix
\bibliographystyle{plainnat} 
\bibliography{references}

\begin{thebibliography}{66}
\providecommand{\natexlab}[1]{#1}
\providecommand{\url}[1]{\texttt{#1}}
\expandafter\ifx\csname urlstyle\endcsname\relax
  \providecommand{\doi}[1]{doi: #1}\else
  \providecommand{\doi}{doi: \begingroup \urlstyle{rm}\Url}\fi

\bibitem[Alnaes et~al.(2015)Alnaes, Blechta, Hake, Johansson, Kehlet, Logg,
  Richardson, Ring, Rognes, and Wells]{AlnaesEtal2015}
M.~S. Alnaes, J.~Blechta, J.~Hake, A.~Johansson, B.~Kehlet, A.~Logg,
  C.~Richardson, J.~Ring, M.~E. Rognes, and G.~N. Wells.
\newblock The {FEniCS} project version 1.5.
\newblock \emph{Archive of Numerical Software}, 3, 2015.
\newblock \doi{10.11588/ans.2015.100.20553}.

\bibitem[Andrieu et~al.(2003)Andrieu, de~Freitas, Doucet, and
  Jordan]{Andrieu2003}
Christophe Andrieu, Nando de~Freitas, Arnaud Doucet, and Michael~I. Jordan.
\newblock {An Introduction to MCMC for Machine Learning}.
\newblock \emph{Machine Learning 2003 50:1}, 50\penalty0 (1):\penalty0 5--43,
  jan 2003.
\newblock ISSN 1573-0565.
\newblock \doi{10.1023/A:1020281327116}.
\newblock URL \url{https://link.springer.com/article/10.1023/A:1020281327116}.

\bibitem[Beck and Teboulle(2009)]{Beck2009FastGA}
A.~Beck and M.~Teboulle.
\newblock Fast gradient-based algorithms for constrained total variation image
  denoising and deblurring problems.
\newblock \emph{IEEE Transactions on Image Processing}, 18:\penalty0
  2419--2434, 2009.

\bibitem[Beskos et~al.(2013)Beskos, Pillai, Roberts, Sanz-Serna, and
  Stuart]{Beskos2013}
Alexandros Beskos, Natesh Pillai, Gareth Roberts, Jesus~Maria Sanz-Serna, and
  Andrew Stuart.
\newblock {Optimal tuning of the hybrid Monte Carlo algorithm}.
\newblock \emph{Bernoulli}, 19\penalty0 (5 A):\penalty0 1501--1534, 2013.
\newblock ISSN 13507265.
\newblock \doi{10.3150/12-BEJ414}.
\newblock URL \url{https://projecteuclid.org/euclid.bj/1383661192}.

\bibitem[Betancourt(2017)]{betancourt2017conceptual}
Michael Betancourt.
\newblock A conceptual introduction to hamiltonian monte carlo.
\newblock \emph{arXiv preprint arXiv:1701.02434}, 2017.

\bibitem[Betancourt et~al.(2017)Betancourt, Byrne, Livingstone, and
  Girolami]{Betancourt2017}
Michael Betancourt, Simon Byrne, Sam Livingstone, and Mark Girolami.
\newblock {The geometric foundations of Hamiltonian Monte Carlo}.
\newblock \emph{https://doi.org/10.3150/16-BEJ810}, 23\penalty0 (4A):\penalty0
  2257--2298, nov 2017.
\newblock ISSN 1350-7265.
\newblock \doi{10.3150/16-BEJ810}.
\newblock URL
  \url{https://projecteuclid.org/journals/bernoulli/volume-23/issue-4A/The-geometric-foundations-of-Hamiltonian-Monte-Carlo/10.3150/16-BEJ810.full
  https://projecteuclid.org/journals/bernoulli/volume-23/issue-4A/The-geometric-foundations-of-Hamiltonian-Monte-Carlo/10.3150/16-BEJ810.short}.

\bibitem[Boomsma et~al.(2013)Boomsma, Frellsen, Harder, Bottaro, Johansson,
  Tian, Stovgaard, Andreetta, Olsson, Valentin, Antonov, Christensen, Borg,
  Jensen, Lindorff-Larsen, Ferkinghoff-Borg, and
  Hamelryck]{Boomsma2013PHAISTOSAF}
W.~Boomsma, J.~Frellsen, T.~Harder, Sandro Bottaro, K.~E. Johansson, Pengfei
  Tian, K.~Stovgaard, Christian Andreetta, S.~Olsson, Jan~B. Valentin, L.~D.
  Antonov, Anders~S. Christensen, M.~Borg, Jan~H. Jensen, K.~Lindorff-Larsen,
  J.~Ferkinghoff-Borg, and T.~Hamelryck.
\newblock Phaistos: A framework for markov chain monte carlo simulation and
  inference of protein structure.
\newblock \emph{Journal of Computational Chemistry}, 34:\penalty0 1697 -- 1705,
  2013.

\bibitem[Brooks et~al.(2011)Brooks, Gelman, Jones, and Meng]{Brooks2011}
Steve Brooks, Andrew Gelman, Galin~L. Jones, and Xiao~Li Meng.
\newblock \emph{{Handbook of Markov Chain Monte Carlo}}.
\newblock CRC Press, may 2011.
\newblock ISBN 9781420079425.
\newblock \doi{10.1201/b10905}.

\bibitem[Carter et~al.(1989)Carter, Ciccotti, Hynes, and
  Kapral]{carter1989constrained}
Emily~A Carter, Giovanni Ciccotti, James~T Hynes, and Raymond Kapral.
\newblock Constrained reaction coordinate dynamics for the simulation of rare
  events.
\newblock \emph{Chemical Physics Letters}, 156\penalty0 (5):\penalty0 472--477,
  1989.

\bibitem[Chaari et~al.(2013)Chaari, Tourneret, and Batatia]{Chaari2013SparseBR}
Lotfi Chaari, J.~Tourneret, and H.~Batatia.
\newblock Sparse bayesian regularization using bernoulli-laplacian priors.
\newblock \emph{21st European Signal Processing Conference (EUSIPCO 2013)},
  pages 1--5, 2013.

\bibitem[Chambolle et~al.(2010)Chambolle, Caselles, Cremers, Novaga, and
  Pock]{Chambolle2010}
Antonin Chambolle, Vicent Caselles, Daniel Cremers, Matteo Novaga, and Thomas
  Pock.
\newblock \emph{{An Introduction to Total Variation for Image Analysis:}},
  pages 263--340.
\newblock De Gruyter, 2010.
\newblock \doi{doi:10.1515/9783110226157.263}.
\newblock URL \url{https://doi.org/10.1515/9783110226157.263}.

\bibitem[Christen and Fox(2005)]{christen2005markov}
J~Andr{\'e}s Christen and Colin Fox.
\newblock Markov chain monte carlo using an approximation.
\newblock \emph{Journal of Computational and Graphical statistics}, 14\penalty0
  (4):\penalty0 795--810, 2005.

\bibitem[Creutz(1988)]{creutz1988global}
Michael Creutz.
\newblock Global monte carlo algorithms for many-fermion systems.
\newblock \emph{Physical Review D}, 38\penalty0 (4):\penalty0 1228, 1988.

\bibitem[Dasgupta et~al.(2024)Dasgupta, Patel, Ray, Johnson, and
  Oberai]{dasgupta2024dimension}
Agnimitra Dasgupta, Dhruv~V Patel, Deep Ray, Erik~A Johnson, and Assad~A
  Oberai.
\newblock A dimension-reduced variational approach for solving physics-based
  inverse problems using generative adversarial network priors and normalizing
  flows.
\newblock \emph{Computer Methods in Applied Mechanics and Engineering},
  420:\penalty0 116682, 2024.

\bibitem[Dashti and Stuart(2017)]{Dashti2017}
Masoumeh Dashti and Andrew~M. Stuart.
\newblock {The Bayesian Approach to Inverse Problems}.
\newblock In \emph{Handbook of Uncertainty Quantification}, pages 311--428.
  Springer, Cham, jun 2017.
\newblock \doi{10.1007/978-3-319-12385-1_7}.
\newblock URL
  \url{https://link.springer.com/referenceworkentry/10.1007/978-3-319-12385-1{\_}7}.

\bibitem[Dauphin et~al.(2014)Dauphin, Pascanu, Gulcehre, Cho, Ganguli, and
  Bengio]{dauphin2014identifying}
Yann~N Dauphin, Razvan Pascanu, Caglar Gulcehre, Kyunghyun Cho, Surya Ganguli,
  and Yoshua Bengio.
\newblock Identifying and attacking the saddle point problem in
  high-dimensional non-convex optimization.
\newblock \emph{Advances in neural information processing systems}, 27, 2014.

\bibitem[Detemple et~al.(2003)Detemple, Garcia, and
  Rindisbacher]{detemple2003monte}
Jerome~B Detemple, Ren Garcia, and Marcel Rindisbacher.
\newblock A monte carlo method for optimal portfolios.
\newblock \emph{The journal of Finance}, 58\penalty0 (1):\penalty0 401--446,
  2003.

\bibitem[Duane et~al.(1987)Duane, Kennedy, Pendleton, and
  Roweth]{duane1987hybrid}
Simon Duane, A.D. Kennedy, Brian~J. Pendleton, and Duncan Roweth.
\newblock Hybrid monte carlo.
\newblock \emph{Physics Letters B}, 195\penalty0 (2):\penalty0 216--222, 1987.
\newblock ISSN 0370-2693.
\newblock \doi{https://doi.org/10.1016/0370-2693(87)91197-X}.

\bibitem[Dyer et~al.(1991)Dyer, Frieze, and Kannan]{dyer1991random}
Martin Dyer, Alan Frieze, and Ravi Kannan.
\newblock A random polynomial-time algorithm for approximating the volume of
  convex bodies.
\newblock \emph{Journal of the ACM (JACM)}, 38\penalty0 (1):\penalty0 1--17,
  1991.

\bibitem[Efendiev et~al.(2006)Efendiev, Hou, and Luo]{Efendiev2006}
Y.~Efendiev, T.~Hou, and W.~Luo.
\newblock {Preconditioning Markov Chain Monte Carlo Simulations Using
  Coarse-Scale Models}.
\newblock \emph{http://dx.doi.org/10.1137/050628568}, 28\penalty0 (2):\penalty0
  776--803, jul 2006.
\newblock ISSN 10648275.
\newblock \doi{10.1137/050628568}.

\bibitem[Farsiu et~al.(1996)Farsiu, Robinson, Elad, and
  Milanfar]{Farsiu1996FastAR}
Sina Farsiu, D.~Robinson, Michael Elad, and P.~Milanfar.
\newblock Fast and robust multi-frame super-resolution.
\newblock \emph{IEEE Transactions on Image Processing}, 1996.

\bibitem[Gelfand and Smith(1990)]{Gelfand1990}
Alan~E. Gelfand and Adrian~F.M. Smith.
\newblock {Sampling-based approaches to calculating marginal densities}.
\newblock \emph{Journal of the American Statistical Association}, 85\penalty0
  (410):\penalty0 398--409, 1990.
\newblock ISSN 1537274X.
\newblock \doi{10.1080/01621459.1990.10476213}.

\bibitem[Gelman et~al.(2014)Gelman, Carlin, Stern, and Rubin]{Gelman2014}
Andrew Gelman, John B~B Carlin, Hal S~S Stern, and Donald B~B Rubin.
\newblock {Bayesian Data Analysis, Third Edition (Texts in Statistical
  Science)}.
\newblock \emph{Book}, page 675, 2014.

\bibitem[Geman and Geman(1984)]{geman1984stochastic}
Stuart Geman and Donald Geman.
\newblock Stochastic relaxation, gibbs distributions, and the bayesian
  restoration of images.
\newblock \emph{IEEE Transactions on pattern analysis and machine
  intelligence}, \penalty0 (6):\penalty0 721--741, 1984.

\bibitem[Geraci et~al.(2017)Geraci, Eldred, and
  Iaccarino]{geraci2017multifidelity}
Gianluca Geraci, Michael~S Eldred, and Gianluca Iaccarino.
\newblock A multifidelity multilevel monte carlo method for uncertainty
  propagation in aerospace applications.
\newblock In \emph{19th AIAA non-deterministic approaches conference}, page
  1951, 2017.

\bibitem[Girolami and Calderhead(2011)]{girolami2011riemann}
Mark Girolami and Ben Calderhead.
\newblock Riemann manifold langevin and hamiltonian monte carlo methods.
\newblock \emph{Journal of the Royal Statistical Society: Series B (Statistical
  Methodology)}, 73\penalty0 (2):\penalty0 123--214, 2011.

\bibitem[Goodfellow et~al.(2014)Goodfellow, Pouget-Abadie, Mirza, Xu,
  Warde-Farley, Ozair, Courville, and Bengio]{Goodfellow2014}
Ian~J. Goodfellow, Jean Pouget-Abadie, Mehdi Mirza, Bing Xu, David
  Warde-Farley, Sherjil Ozair, Aaron Courville, and Yoshua Bengio.
\newblock {Generative Adversarial Networks}.
\newblock jun 2014.
\newblock URL \url{http://arxiv.org/abs/1406.2661}.

\bibitem[Haario et~al.(2001)Haario, Saksman, and Tamminen]{Haario2001}
Heikki Haario, Eero Saksman, and Johanna Tamminen.
\newblock {An adaptive Metropolis algorithm}.
\newblock \emph{Bernoulli}, 7\penalty0 (2):\penalty0 223 -- 242, 2001.

\bibitem[Habeck et~al.(2005)Habeck, Nilges, and Rieping]{habeck2005replica}
Michael Habeck, Michael Nilges, and Wolfgang Rieping.
\newblock Replica-exchange monte carlo scheme for bayesian data analysis.
\newblock \emph{Physical review letters}, 94\penalty0 (1):\penalty0 018105,
  2005.

\bibitem[Hammond et~al.(2014)Hammond, Lichtner, and Mills]{pflotran-paper}
Glenn~E. Hammond, Peter~C. Lichtner, and Richard~T. Mills.
\newblock Evaluating the performance of parallel subsurface simulators: An
  illustrative example with pflotran.
\newblock \emph{Water Resources Research}, 50:\penalty0 208--228, 2014.
\newblock \doi{10.1002/2012WR013483}.

\bibitem[Harbaugh(2005)]{harbaugh2005modflow}
Arlen~W Harbaugh.
\newblock \emph{MODFLOW-2005, the US Geological Survey modular ground-water
  model: the ground-water flow process}, volume~6.
\newblock US Department of the Interior, US Geological Survey Reston, VA, USA,
  2005.

\bibitem[Ho et~al.(2020)Ho, Jain, and Abbeel]{ho2020denoising}
Jonathan Ho, Ajay Jain, and Pieter Abbeel.
\newblock Denoising diffusion probabilistic models.
\newblock \emph{Advances in neural information processing systems},
  33:\penalty0 6840--6851, 2020.

\bibitem[Hoffman et~al.(2021)Hoffman, Radul, and Sountsov]{hoffman2021adaptive}
Matthew Hoffman, Alexey Radul, and Pavel Sountsov.
\newblock An adaptive-mcmc scheme for setting trajectory lengths in hamiltonian
  monte carlo.
\newblock In Arindam Banerjee and Kenji Fukumizu, editors, \emph{Proceedings of
  The 24th International Conference on Artificial Intelligence and Statistics},
  volume 130 of \emph{Proceedings of Machine Learning Research}, pages
  3907--3915. PMLR, 13--15 Apr 2021.
\newblock URL \url{https://proceedings.mlr.press/v130/hoffman21a.html}.

\bibitem[Hoffman et~al.(2014)Hoffman, Gelman, et~al.]{hoffman2014no}
Matthew~D Hoffman, Andrew Gelman, et~al.
\newblock The no-u-turn sampler: adaptively setting path lengths in hamiltonian
  monte carlo.
\newblock \emph{J. Mach. Learn. Res.}, 15\penalty0 (1):\penalty0 1593--1623,
  2014.

\bibitem[Iglesias et~al.(2013)Iglesias, Law, and
  Stuart]{Iglesias2013EvaluationOG}
M~Iglesias, K~Law, and A~Stuart.
\newblock {Evaluation of Gaussian approximations for data assimilation in
  reservoir models}.
\newblock \emph{Computational Geosciences}, 17:\penalty0 851--885, 2013.

\bibitem[Illman et~al.(2007)Illman, Liu, and Craig]{illman2007steady}
Walter~A Illman, Xiaoyi Liu, and Andrew Craig.
\newblock Steady-state hydraulic tomography in a laboratory aquifer with
  deterministic heterogeneity: Multi-method and multiscale validation of
  hydraulic conductivity tomograms.
\newblock \emph{Journal of Hydrology}, 341\penalty0 (3-4):\penalty0 222--234,
  2007.

\bibitem[JJ(2004)]{JJFern2004}
Fern{\'{a}}ndez-Dur{\'{a}}n JJ.
\newblock {Circular distributions based on nonnegative trigonometric sums}.
\newblock \emph{Biometrics}, 60\penalty0 (2):\penalty0 499--503, jun 2004.
\newblock ISSN 0006-341X.
\newblock \doi{10.1111/J.0006-341X.2004.00195.X}.
\newblock URL \url{https://pubmed.ncbi.nlm.nih.gov/15180676/}.

\bibitem[Kaipio and Somersalo(2006)]{kaipio2006statistical}
Jari Kaipio and Erkki Somersalo.
\newblock \emph{Statistical and computational inverse problems}, volume 160.
\newblock Springer Science \& Business Media, 2006.

\bibitem[Kaipio and Fox(2011)]{Kaipio2011}
Jari~P Kaipio and Colin Fox.
\newblock {The Bayesian Framework for Inverse Problems in Heat Transfer}.
\newblock \emph{Heat Transfer Engineering}, 32\penalty0 (9):\penalty0 718--753,
  2011.
\newblock \doi{10.1080/01457632.2011.525137}.
\newblock URL \url{https://doi.org/10.1080/01457632.2011.525137}.

\bibitem[Keyes et~al.(2013)Keyes, McInnes, Woodward, Gropp, Myra, Pernice,
  Bell, Brown, Clo, Connors, Constantinescu, Estep, Evans, Farhat, Hakim,
  Hammond, Hansen, Hill, Isaac, Jiao, Jordan, Kaushik, Kaxiras, Koniges, Lee,
  Lott, Lu, Magerlein, Maxwell, McCourt, Mehl, Pawlowski, Randles, Reynolds,
  Rivière, Rüde, Scheibe, Shadid, Sheehan, Shephard, Siegel, Smith, Tang,
  Wilson, and Wohlmuth]{Keyes_etal_13}
David~E Keyes, Lois~C McInnes, Carol Woodward, William Gropp, Eric Myra,
  Michael Pernice, John Bell, Jed Brown, Alain Clo, Jeffrey Connors, Emil
  Constantinescu, Don Estep, Kate Evans, Charbel Farhat, Ammar Hakim, Glenn
  Hammond, Glen Hansen, Judith Hill, Tobin Isaac, Xiangmin Jiao, Kirk Jordan,
  Dinesh Kaushik, Efthimios Kaxiras, Alice Koniges, Kihwan Lee, Aaron Lott,
  Qiming Lu, John Magerlein, Reed Maxwell, Michael McCourt, Miriam Mehl, Roger
  Pawlowski, Amanda~P Randles, Daniel Reynolds, Beatrice Rivière, Ulrich
  Rüde, Tim Scheibe, John Shadid, Brendan Sheehan, Mark Shephard, Andrew
  Siegel, Barry Smith, Xianzhu Tang, Cian Wilson, and Barbara Wohlmuth.
\newblock Multiphysics simulations: Challenges and opportunities.
\newblock \emph{The International Journal of High Performance Computing
  Applications}, 27\penalty0 (1):\penalty0 4--83, 2013.
\newblock \doi{10.1177/1094342012468181}.

\bibitem[Kingma and Ba(2014)]{Kingma2014}
Diederik~P. Kingma and Jimmy Ba.
\newblock {Adam: A Method for Stochastic Optimization}.
\newblock dec 2014.
\newblock URL \url{http://arxiv.org/abs/1412.6980}.

\bibitem[Koller and Friedman(2009)]{koller2009probabilistic}
Daphne Koller and Nir Friedman.
\newblock \emph{Probabilistic graphical models: principles and techniques}.
\newblock MIT press, 2009.

\bibitem[Lee and Kitanidis(2013)]{lee2013bayesian}
J~Lee and PK~Kitanidis.
\newblock Bayesian inversion with total variation prior for discrete geologic
  structure identification.
\newblock \emph{Water Resources Research}, 49\penalty0 (11):\penalty0
  7658--7669, 2013.

\bibitem[Li et~al.(2020)Li, Kovachki, Azizzadenesheli, Liu, Bhattacharya,
  Stuart, and Anandkumar]{li2020fourier}
Zongyi Li, Nikola Kovachki, Kamyar Azizzadenesheli, Burigede Liu, Kaushik
  Bhattacharya, Andrew Stuart, and Anima Anandkumar.
\newblock Fourier neural operator for parametric partial differential
  equations.
\newblock \emph{arXiv preprint arXiv:2010.08895}, 2020.

\bibitem[Liu and Kitanidis(2011)]{Liu2011}
X.~Liu and P.~K. Kitanidis.
\newblock {Large-scale inverse modeling with an application in hydraulic
  tomography}.
\newblock \emph{Water Resources Research}, 47\penalty0 (2):\penalty0 2501, feb
  2011.
\newblock ISSN 1944-7973.
\newblock \doi{10.1029/2010WR009144}.
\newblock URL
  \url{https://onlinelibrary-wiley-com.stanford.idm.oclc.org/doi/full/10.1029/2010WR009144
  https://onlinelibrary-wiley-com.stanford.idm.oclc.org/doi/abs/10.1029/2010WR009144
  https://agupubs-onlinelibrary-wiley-com.stanford.idm.oclc.org/doi/10.1029/2010WR009144}.

\bibitem[Lu et~al.(2021)Lu, Jin, Pang, Zhang, and Karniadakis]{lu2021learning}
Lu~Lu, Pengzhan Jin, Guofei Pang, Zhongqiang Zhang, and George~Em Karniadakis.
\newblock Learning nonlinear operators via deeponet based on the universal
  approximation theorem of operators.
\newblock \emph{Nature Machine Intelligence}, 3\penalty0 (3):\penalty0
  218--229, 2021.

\bibitem[Martin et~al.(2012)Martin, Wilcox, Burstedde, and Ghattas]{Martin2012}
James Martin, Lucas~C. Wilcox, Carsten Burstedde, and Omar Ghattas.
\newblock {A Stochastic Newton MCMC Method for Large-Scale Statistical Inverse
  Problems with Application to Seismic Inversion}.
\newblock \emph{SIAM Journal on Scientific Computing}, 34\penalty0
  (3):\penalty0 A1460--A1487, jan 2012.
\newblock ISSN 1064-8275.
\newblock \doi{10.1137/110845598}.
\newblock URL \url{http://epubs.siam.org/doi/10.1137/110845598}.

\bibitem[Metropolis et~al.(1953)Metropolis, Rosenbluth, Rosenbluth, Teller, and
  Teller]{Metropolis1953}
Nicholas Metropolis, Arianna~W. Rosenbluth, Marshall~N. Rosenbluth, Augusta~H.
  Teller, and Edward Teller.
\newblock Equation of state calculations by fast computing machines.
\newblock \emph{The Journal of Chemical Physics}, 21:\penalty0 1087, 12 1953.
\newblock ISSN 0021-9606.
\newblock \doi{10.1063/1.1699114}.
\newblock URL \url{https://aip.scitation.org/doi/abs/10.1063/1.1699114}.

\bibitem[Mosser et~al.(2019)Mosser, Dubrule, and Blunt]{Mosser2019}
Lukas Mosser, Olivier Dubrule, and Martin~J. Blunt.
\newblock {Stochastic Seismic Waveform Inversion Using Generative Adversarial
  Networks as a Geological Prior}.
\newblock \emph{Mathematical Geosciences 2019 52:1}, 52\penalty0 (1):\penalty0
  53--79, nov 2019.
\newblock ISSN 1874-8953.
\newblock \doi{10.1007/S11004-019-09832-6}.
\newblock URL
  \url{https://link.springer.com/article/10.1007/s11004-019-09832-6}.

\bibitem[Neal(2012)]{Neal2012}
Radford~M. Neal.
\newblock {MCMC using Hamiltonian dynamics}.
\newblock \emph{Handbook of Markov Chain Monte Carlo}, pages 1--592, jun 2012.
\newblock \doi{10.1201/b10905}.
\newblock URL \url{http://arxiv.org/abs/1206.1901
  http://dx.doi.org/10.1201/b10905}.

\bibitem[Nijkamp et~al.(2019)Nijkamp, Hill, Han, Zhu, and Wu]{Nijkamp2019}
Erik Nijkamp, Mitch Hill, Tian Han, Song~Chun Zhu, and Ying~Nian Wu.
\newblock {On the Anatomy of MCMC-Based Maximum Likelihood Learning of
  Energy-Based Models}.
\newblock \emph{AAAI 2020 - 34th AAAI Conference on Artificial Intelligence},
  pages 5272--5280, mar 2019.
\newblock ISSN 2159-5399.
\newblock \doi{10.1609/aaai.v34i04.5973}.
\newblock URL \url{https://arxiv.org/abs/1903.12370v4}.

\bibitem[Park and Casella(2012)]{Park2012}
Trevor Park and George Casella.
\newblock {The Bayesian Lasso}.
\newblock \emph{https://doi.org/10.1198/016214508000000337}, 103\penalty0
  (482):\penalty0 681--686, jun 2012.
\newblock \doi{10.1198/016214508000000337}.
\newblock URL
  \url{https://www.tandfonline.com/doi/abs/10.1198/016214508000000337}.

\bibitem[Parno and Marzouk(2018)]{parno2018transport}
Matthew~D Parno and Youssef~M Marzouk.
\newblock Transport map accelerated markov chain monte carlo.
\newblock \emph{SIAM/ASA Journal on Uncertainty Quantification}, 6\penalty0
  (2):\penalty0 645--682, 2018.

\bibitem[Patel and Oberai(2019)]{patel2019bayesian}
Dhruv Patel and Assad~A Oberai.
\newblock Bayesian inference with generative adversarial network priors.
\newblock \emph{arXiv preprint arXiv:1907.09987}, 2019.

\bibitem[Patel et~al.(2024)Patel, Ray, Abdelmalik, Hughes, and
  Oberai]{patel2024variationally}
Dhruv Patel, Deep Ray, Michael~RA Abdelmalik, Thomas~JR Hughes, and Assad~A
  Oberai.
\newblock Variationally mimetic operator networks.
\newblock \emph{Computer Methods in Applied Mechanics and Engineering},
  419:\penalty0 116536, 2024.

\bibitem[Patel and Oberai(2021)]{Patel2021GAN}
Dhruv~V. Patel and Assad~A. Oberai.
\newblock {GAN-Based Priors for Quantifying Uncertainty in Supervised
  Learning}.
\newblock \emph{SIAM/ASA Journal on Uncertainty Quantification}, 9\penalty0
  (3):\penalty0 1314--1343, jan 2021.
\newblock \doi{10.1137/20M1354210}.

\bibitem[Patel et~al.(2022)Patel, Ray, and Oberai]{patel2022solution}
Dhruv~V Patel, Deep Ray, and Assad~A Oberai.
\newblock Solution of physics-based bayesian inverse problems with deep
  generative priors.
\newblock \emph{Computer Methods in Applied Mechanics and Engineering},
  400:\penalty0 115428, 2022.

\bibitem[Plessix(2006)]{plessix2006}
R.-E. Plessix.
\newblock {A review of the adjoint-state method for computing the gradient of a
  functional with geophysical applications}.
\newblock \emph{Geophysical Journal International}, 167\penalty0 (2):\penalty0
  495--503, 11 2006.
\newblock ISSN 0956-540X.
\newblock \doi{10.1111/j.1365-246X.2006.02978.x}.
\newblock URL \url{https://doi.org/10.1111/j.1365-246X.2006.02978.x}.

\bibitem[Roberts and Tweedie(1996)]{roberts1996exponential}
Gareth~O Roberts and Richard~L Tweedie.
\newblock Exponential convergence of langevin distributions and their discrete
  approximations.
\newblock \emph{Bernoulli}, pages 341--363, 1996.

\bibitem[Rosso et~al.(2002)Rosso, Min{\'a}ry, Zhu, and Tuckerman]{rosso2002use}
Lula Rosso, Peter Min{\'a}ry, Zhongwei Zhu, and Mark~E Tuckerman.
\newblock On the use of the adiabatic molecular dynamics technique in the
  calculation of free energy profiles.
\newblock \emph{The Journal of chemical physics}, 116\penalty0 (11):\penalty0
  4389--4402, 2002.

\bibitem[Rudin et~al.(1992)Rudin, Osher, and Fatemi]{Rudin1992}
Leonid~I. Rudin, Stanley Osher, and Emad Fatemi.
\newblock {Nonlinear total variation based noise removal algorithms}.
\newblock \emph{Physica D: Nonlinear Phenomena}, 60\penalty0 (1-4):\penalty0
  259--268, nov 1992.
\newblock ISSN 0167-2789.
\newblock \doi{10.1016/0167-2789(92)90242-F}.

\bibitem[Sherlock et~al.(2017)Sherlock, Golightly, and
  Henderson]{sherlock2017adaptive}
Chris Sherlock, Andrew Golightly, and Daniel~A Henderson.
\newblock Adaptive, delayed-acceptance mcmc for targets with expensive
  likelihoods.
\newblock \emph{Journal of Computational and Graphical Statistics}, 26\penalty0
  (2):\penalty0 434--444, 2017.

\bibitem[Vauhkonen et~al.(1997)Vauhkonen, Kaipio, Somersalo, and
  Karjalainen]{Vauhkonen1997}
M~Vauhkonen, J~P Kaipio, E~Somersalo, and P~A Karjalainen.
\newblock {Electrical impedance tomography with basis constraints}.
\newblock \emph{Inverse Problems}, 13\penalty0 (2):\penalty0 523--530, apr
  1997.
\newblock ISSN 0266-5611.
\newblock \doi{10.1088/0266-5611/13/2/020}.
\newblock URL
  \url{http://stacks.iop.org/0266-5611/13/i=2/a=020?key=crossref.46559bf45aab26a8302acc14e8db4c89}.

\bibitem[Welling and Teh(2011)]{welling2011bayesian}
Max Welling and Yee~W Teh.
\newblock Bayesian learning via stochastic gradient langevin dynamics.
\newblock In \emph{Proceedings of the 28th international conference on machine
  learning (ICML-11)}, pages 681--688. Citeseer, 2011.

\bibitem[Zheng et~al.(2013)Zheng, Rohrdanz, and Clementi]{zheng2013rapid}
Wenwei Zheng, Mary~A Rohrdanz, and Cecilia Clementi.
\newblock Rapid exploration of configuration space with diffusion-map-directed
  molecular dynamics.
\newblock \emph{The journal of physical chemistry B}, 117\penalty0
  (42):\penalty0 12769--12776, 2013.

\bibitem[Zhu et~al.(2019)Zhu, Zabaras, Koutsourelakis, and Perdikaris]{Zhu2019}
Yinhao Zhu, Nicholas Zabaras, Phaedon~Stelios Koutsourelakis, and Paris
  Perdikaris.
\newblock {Physics-constrained deep learning for high-dimensional surrogate
  modeling and uncertainty quantification without labeled data}.
\newblock \emph{Journal of Computational Physics}, 394:\penalty0 56--81, oct
  2019.
\newblock ISSN 0021-9991.
\newblock \doi{10.1016/J.JCP.2019.05.024}.

\end{thebibliography}

\newpage
\appendix
\section{Details of the model architectures}
In this section we describe the architectures of different deep learning models (used as a prior or surrogate forward models) for various inverse problems. Some of the nomenclature we use are as follows:
\begin{enumerate}
		\item FC($n$) --- Fully connected layer of width $n$.
  
            \item LReLU($\alpha$), ReLU TanH --- Leaky rectified linear unit (with negative slope parameter $\alpha$), rectified linear unit, and hyperbolic tangent activation functions, respectively.
            
  		\item BN --- batch normalization.
    
            \item Conv2D ($c_{\mathrm{out}}$, $k$, $s$, $p$) --- 2D convolution layer $c_{\mathrm{out}}$ output channels, kernel size $(k,k)$, stride $s$ and padding $p$.
            
            \item Conv2D ($c_{\mathrm{out}}$, ($k_v$, $k_h$), $s$, $p$) --- 2D convolution layer $c_{\mathrm{out}}$ output channels, kernel size $(k_v, k_h)$, stride $s$ and padding $p$.       
		\item Tr. Conv2D ($c_{\mathrm{out}}$, $k$, $s$, $p$) --- 2D transpose convolution layer with $c_{\mathrm{out}}$ output channels, kernel size $(k,k)$, stride $s$, padding $p$.

            \item Tr. Conv2D ($c_{\mathrm{out}}$, ($k_v, k_h$), $s$, $p$) --- 2D transpose convolution layer with $c_{\mathrm{out}}$ output channels, kernel size $(k_v, k_h)$, stride $s$, padding $p$.
\end{enumerate}

\begin{figure}[htbp]
    \centering
    \includegraphics[width=0.45\linewidth]{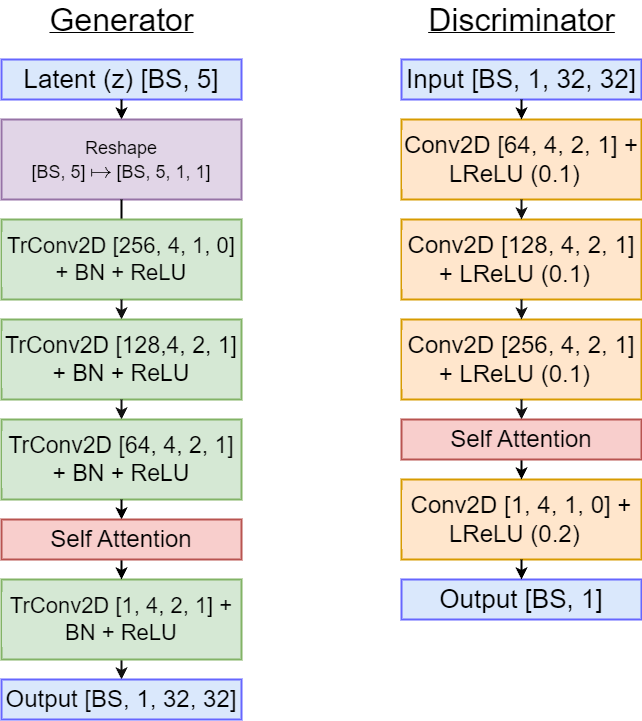}
    \caption{Architectures for the initial condition inversion problem (\Cref{sec:init_cond_inv_gan_prior})}
    \label{fig:arch_init_cond}
\end{figure}
\begin{figure}[htbp]
    \centering
    \includegraphics[width=0.7\linewidth]{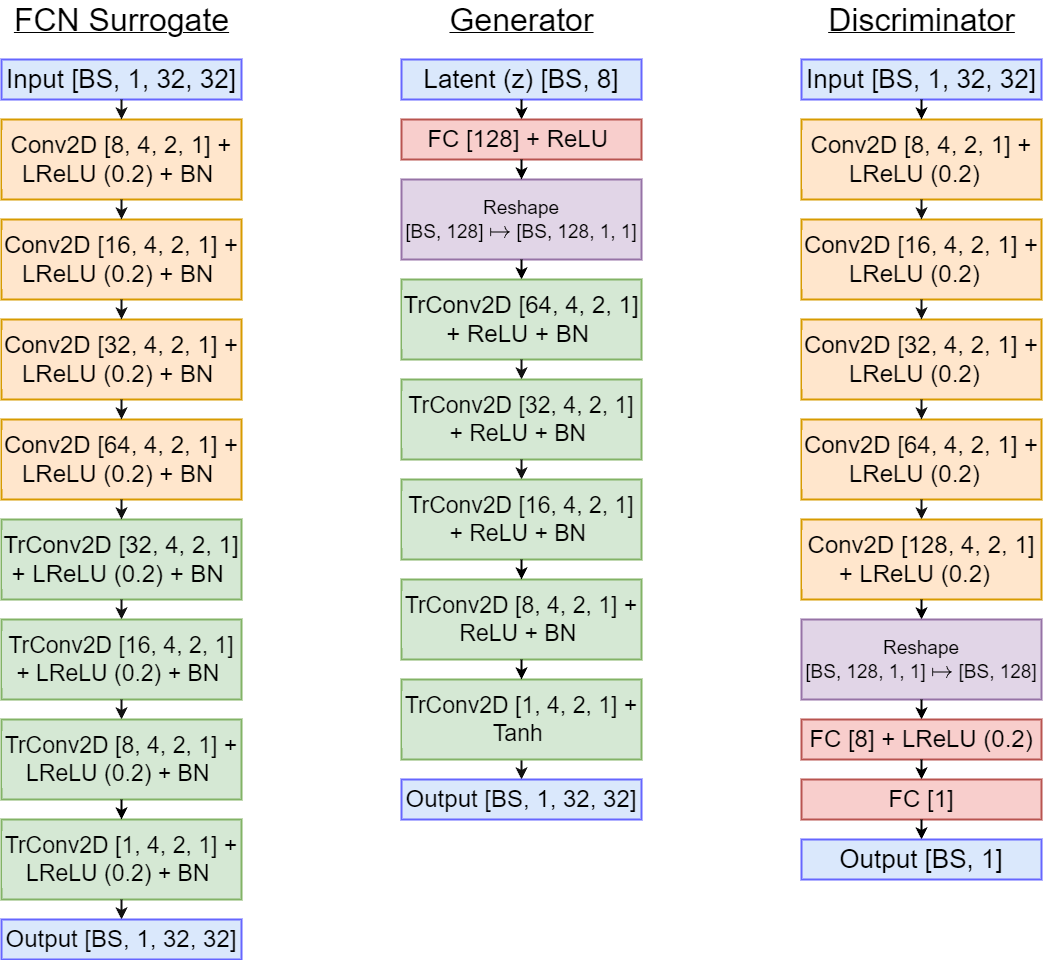}
    \caption{Architectures for the coefficient inversion problem (Section \ref{sec:coefficient_inv})}
    \label{fig:arch_channelized_flow}
\end{figure}
\begin{figure}[htbp]
    \centering
    \includegraphics[width=0.7\linewidth]{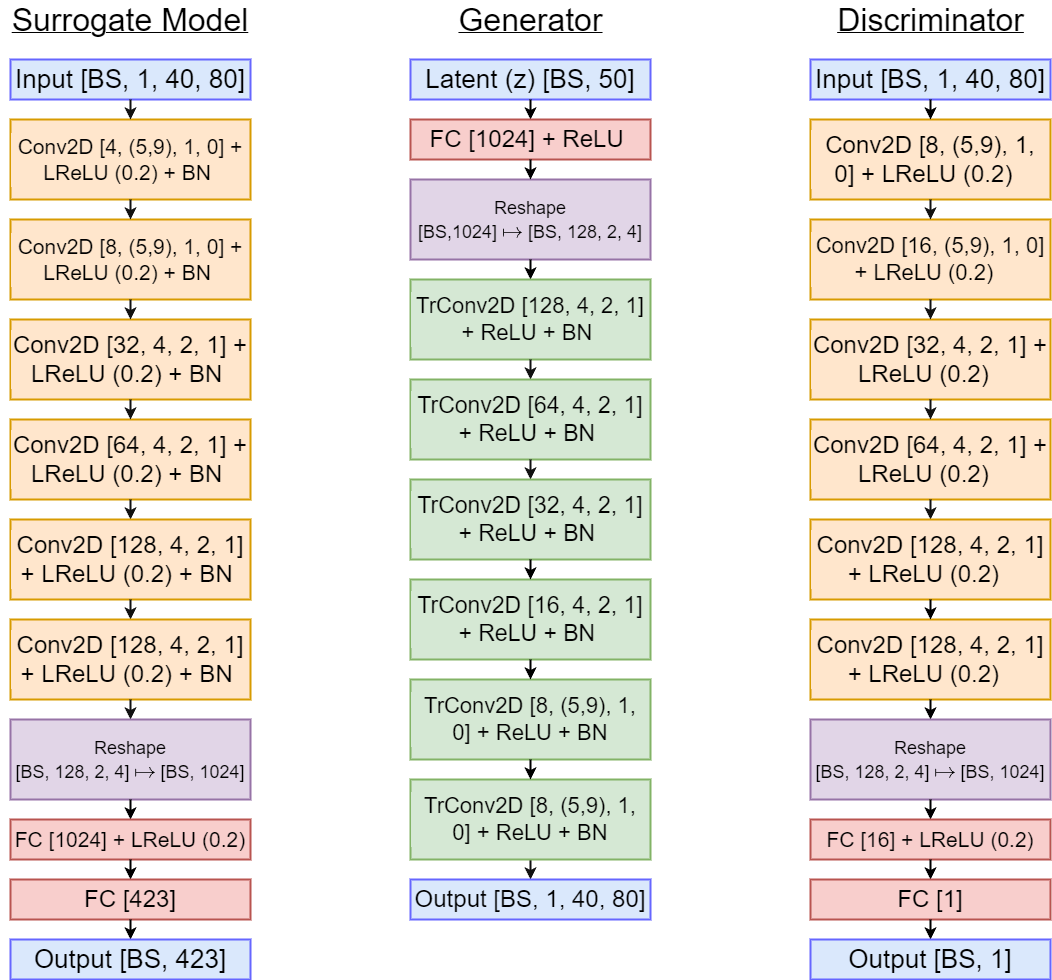}
    \caption{Architectures for the hydraulic tomography problem (Section \ref{sec:real_tomography})}
    \label{fig:arch_real_tomo}
\end{figure}

\end{document}